\documentclass[reprint,aps,prb,superscriptaddress,amsmath,amssymb,floatfix,footinbib,longbibliography]
{revtex4-1}
\usepackage[a4paper,total={6.8in,10in}]{geometry}
\usepackage{amsmath}
\usepackage{mathtools}
\usepackage{graphicx}
\usepackage{xcolor}
\usepackage{physics}
\usepackage[colorlinks, linkcolor= blue, citecolor = blue, urlcolor=blue]{hyperref}
\usepackage[none]{hyphenat}
\usepackage{bm}
\usepackage{comment}
\usepackage{wrapfig}
\usepackage{physics}
\usepackage{lmodern}
\usepackage{dcolumn}
\usepackage{natbib}

\usepackage{soul}

\def\nn{\nonumber}
\def\bea{\begin{eqnarray}}
\def\eea{\end{eqnarray}}
\def\be{\begin{equation}}
\def\ee{\end{equation}}

\def\tc{\textcolor}

\def\kb{\boldsymbol{k}}

\def\e{\varepsilon}
\def\m{\mathcal}
\def\bal{\begin{aligned}}
\def\eal{\end{aligned}}

\begin{document}
\title{Quantum kinetic theory of nonlinear thermal current}
\author{Harsh Varshney}
\email{hvarshny@iitk.ac.in}
\affiliation{Department of Physics, Indian Institute of Technology, Kanpur-208016, India.}
\author{Kamal Das}
\email{kamaldas@iitk.ac.in}
\affiliation{Department of Physics, Indian Institute of Technology, Kanpur-208016, India.}
\author{Pankaj Bhalla}
\email{pankaj.b@srmap.edu.in}
\affiliation{Department of Physics, School of Engineering and Sciences, SRM University AP, Amaravati, 522240, India}
\author{Amit Agarwal}
\email{amitag@iitk.ac.in}
\affiliation{Department of Physics, Indian Institute of Technology, Kanpur-208016, India.}
\begin{abstract}
We investigate the second-order nonlinear electronic thermal  transport induced by temperature gradient. We develop the quantum kinetic theory framework to describe thermal transport in the presence of a temperature gradient. Using this, we predict an intrinsic scattering time-independent nonlinear thermal current in addition to the known extrinsic nonlinear Drude and Berry curvature dipole contributions. We show that the intrinsic thermal current is determined by the band geometric quantities and is non-zero only in systems where both the space inversion and time-reversal symmetries are broken. We employ the developed theory to study the thermal response in tilted massive Dirac systems. We show that besides the different scattering time dependents, the various current contributions have distinct temperature dependencies in the low-temperature limit. Our systematic and comprehensive theory for nonlinear thermal transport paves the way for future theoretical and experimental studies on intrinsic thermal responses.
\end{abstract}

\maketitle

\section{Introduction}

The temperature gradient in a system governs several nontrivial electronic transport phenomena~\cite{SSP,chernodub_PR20221_thermal,nieh_PRB2020}. Particularly,  the anomalous Hall-type effects, where a transverse response is generated due to longitudinal temperature gradient in the absence of a magnetic field, are very fascinating~\cite{Qian_Nu_PRL_2006, Lifa_Zhang_NJP2016}. Example of such Hall-type effect includes the Nernst effect~\cite{Qian_Nu_PRL_2006} that causes a transverse charge flow and the Righi-Leduc (thermal Hall) effect~\cite{matsumoto_PRB2011, yago17} which causes a Hall-type heat flow. These effects, which can be thought of as the thermoelectric and thermal generalization of the charge Hall effect due to the electric field, are primarily investigated in the linear regime where the Hall current (or voltage) is linearly proportional to the temperature gradient. The prerequisite for realizing these effects within the linear response regime is that the time-reversal symmetry of the system must be broken. Otherwise, the Hall responses disappear following Onsager's relation~\cite{Book1}. 

Recently, it has been demonstrated that in the second-order nonlinear (NL) response regime, we can observe the anomalous Hall effects even in systems with time-reversal symmetry~\cite{Sodemann2015, Sinha2022}. The Berry curvature dipole in non-centrosymmetric materials with low crystalline symmetries has been shown to generate the NL Nernst~\cite{Xiao_PRB_2019,wang_PRB2022} as well as NL thermal Hall effect~\cite{Nandy20,bhalla_PRB2021, wang_PRB2022}, both of which are extrinsic (scattering time-dependent) in nature. Interestingly, these Berry curvature dipole-induced Hall effects can also probe the topological phase-transition of valley-Chern type~\cite{Sinha2022,chakraborty_2DM2022_non}. These recent theoretical developments with the experimental realizations of this new class of Hall effects have generated significant interest in NL transport.
%
\begin{figure}[t]
    \centering
    \includegraphics[width = \linewidth]{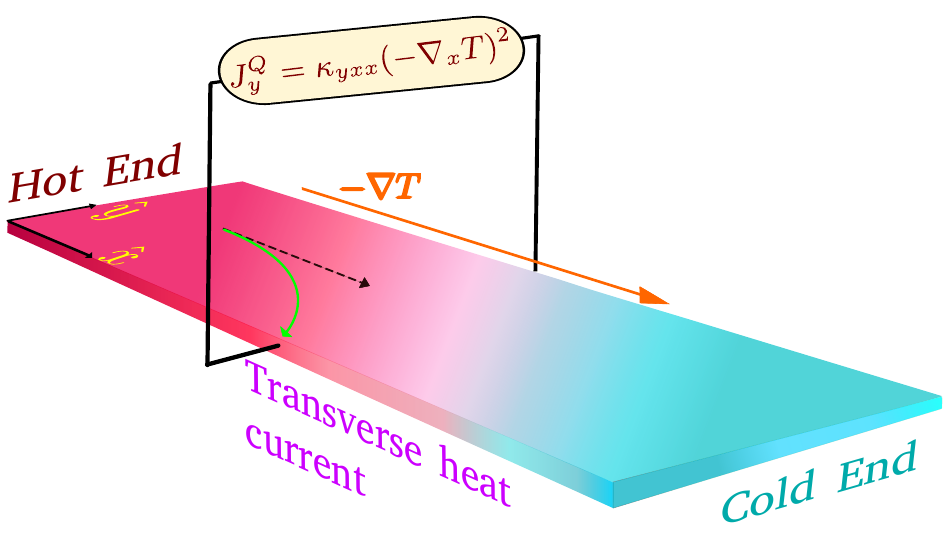}
    \caption{Schematic experimental
setup for measuring the nonlinear
anomalous thermal Hall effect. Here, the dark pink color shows the hot end of the sample, while the light blue color shows the cold end. This sets a temperature gradient ($-\nabla T$) along the $x$ direction. We probe the nonlinear thermal Hall response ($J^{Q}$) along the $y$ (Hall) and the $x$ (longitudinal) direction.}
    \label{fig:fig1}
\end{figure}
%
More recently, an intrinsic scattering time-independent NL Hall effect induced by an electric field has been demonstrated in systems with broken inversion and time-reversal symmetry~\cite{wang_PRL2021_intrin,liu_PRL2021_intrin, Bhalla_shg_2022,shibalik_intrinsic_2022}. The Berry connection polarizability, a band geometric quantity, causes this intrinsic NL Hall effect.  Intriguingly, a similar intrinsic NL Hall effect has also been shown to exist in thermoelectric transport, and it originates from the orbital quadrupole magnetic moment~\cite{gao_PRB2018}. Dissipationless intrinsic Hall effects are very significant as they are solely determined by the band geometric properties and independent of the details of scattering mechanisms. 

Motivated by these recent developments, here we investigate the second-order intrinsic thermal Hall effect using the quantum kinetic theory of thermal transport. Extending the theory developed in Ref.~[\onlinecite{Nagaosa20}], we construct a definition of NL thermal current induced by a  temperature gradient. Using this definition, we calculate all the possible contributions to the NL thermal current. We find an intrinsic NL thermal current that originates from the band geometric quantities of the system. Using symmetry analysis, we show that both the time-reversal symmetry and inversion symmetry of a system must be broken to observe the intrinsic second-order NL thermal current. We show that the parity-time reversal symmetric systems (bipartite antiferromagnets) are good candidates to explore the intrinsic NL thermal Hall effects. In addition, we also obtain the thermal counterpart of the NL Drude (NLD) current and the thermal counterpart of the Berry curvature dipole current. 
Our work provides the realization of all three of these currents on the same footing using the quantum kinetic theory for the first time. We use the developed framework to study the NL thermal Hall transport in tilted massive Dirac systems. 
The extrinsic contributions of the NL thermal current were recently also explored in Ref.~[\onlinecite{wang_PRB2022}]. Here,  we primarily focus on the intrinsic contribution to the NL thermal current. {Our predictions for the non-linear responses become the dominant contribution in systems that preserve time-reversal symmetry or in which the linear response is negligible.} 

This paper is organized as follows. In Sec.~\ref{sec:SecII} we discuss the quantum kinetic framework for the NL thermal current. We calculate the non-equilibrium density matrix and use it to obtain the intrinsic and extrinsic contributions of the thermal current. In Sec.~\ref{sec:SecIII}, we study the longitudinal and transverse NL thermal current in a tilted  massive Dirac system. We summarize our findings in Sec.~\ref{sec:SecIV}.

\section{Quantum kinetic theory for thermal current}
\label{sec:SecII}

We use the quantum kinetic equation to evaluate the density matrix in the crystal momentum representation $\rho(\kb,t)$. Combining it with the relaxation time approximation to account for scattering events, we have \cite{Nagaosa20, Culcer_PRB_2017, Liboff__2006}, 
\bea \label{QKL_eqn}
\frac{\partial \rho(\kb,t)}{\partial t} +\frac{i}{\hbar}\left[ \m{H}_0,\rho(\kb,t)\right] + \frac{\rho(\kb,t)}{\tau}= D_T [\rho(\kb,t)]. \nn \\ 
\eea
%
Here, $\m{H}_0$ represents the grand canonical Hamiltonian which satisfies  $\m{H}_0 \vert u^n_{\kb}\rangle = \tilde{\e}_n \vert u^n_{\kb}\rangle$ with $\tilde{\e}_n = (\varepsilon_{n} - \mu)$. The chemical potential is denoted by $\mu$ and $\e_n $ and $\vert u^n_{\kb}\rangle $ are the $n$-th  energy band and eigenstate of the corresponding Bloch Hamiltonian. 
In Eq.~\eqref{QKL_eqn}, $\tau$ is the relaxation time which we consider to be a constant in our work, and $[\cdot,\cdot]$ denotes the commutator bracket. 
For simplicity of notation, we express $\rho(\kb,t)$ as $\rho$ throughout our calculation. 
In Eq.~\eqref{QKL_eqn}, the thermal driving term \cite{Nagaosa20} is defined as  $D_T (\rho)= - \frac{1}{2\hbar}\bm{E}_T\cdot \left[ \lbrace \m{H}_0 , \pdv{\rho}{\kb} \rbrace - i[\bm{\m{R}_{\kb}},\lbrace \m{H}_0 , \rho \rbrace] \right] $, where the bracket $\lbrace \cdot, \cdot \rbrace$ represents the anticommutator and $\bm{E}_T \equiv -\bm{\nabla}T/ T  $ is the thermal field~\cite{tatara} with $T$ being the temperature. The quantity ${\bm {\m R}_{\kb}}$ in Eq.~\eqref{QKL_eqn} is the momentum space Berry connection. The band-resolved momentum space Berry curvature~\cite{xiao_RMP2010} is defined as ${\bm {\m R}}_{np} = i\langle u^n_{\kb} \vert \partial_{\kb} \vert u^p_{\kb} \rangle$.  

The main challenge for calculating the thermal conductivities within the quantum kinetic theory framework is to construct the definition of the NL heat current. In the semiclassical framework, the transport heat current is obtained by subtracting the immeasurable magnetization heat current from the local heat current~\cite{Qian_Nu_PRL_2006}. In the quantum kinetic theory, we follow the same philosophy. We construct the definition of linear heat current  as~\cite{Nagaosa20}  
{
\bea\label{Th_C_def}
\bm{J} & = &{\rm{Tr}}\bigg[\frac{1}{2}\{\m{H}_0,\bm{v} \}\rho_{\rm D}^{(1)}\bigg] + {\rm{Tr}}\bigg[\frac{1}{2}\{\m{H}_0,\bm{v} \}\rho_{\rm O}^{(1)}\bigg] \nn \\ 
&& + {\rm{Tr}}\left[(\bm{E}_T \times \bm{m}_N) \m{H}_0 \rho_0 \right] + 2{\rm Tr}[\bm{E}_T \times \bm{M}_{\bm \Omega}] ~.
\eea%
}Here, $\rho_{\rm D} (\rho_{\rm O})$ represents the diagonal (off-diagonal) part of the density matrix. The first two terms give the contribution from the heat current operator $\frac{1}{2}\{ {\mathcal H}_0, {\bm v}\}$ with $\bm{v}$ and $\rho^{(1)} = \rho^{(1)}_{\rm D}  + \rho^{(1)}_{\rm O} $ as the velocity operator and first-order density matrix. The third term is the contribution from the particle magnetic moment $\bm{m}_N \equiv \bm{m}/-e$, with $\bm{m}$ being the orbital magnetic moment (OMM) of the Bloch electrons and $-e$ being the electric charge of the electrons. Similar to the semiclassical framework, it has been added to cancel the immeasurable OMM-related term originating from the heat current operator term [second term in Eq.~\eqref{Th_C_def}]. The fourth term in Eq.~\eqref{Th_C_def} stands for the thermal current due to the Berry curvature-induced heat magnetisation~\cite{Qian_PRL_2011,Sumiyoshi_JPS_2013,Shitade_2014,Andrey_PRL_2015,Nakai_JOP_2016}. The Berry curvature-induced heat magnetization is given by~\cite{Lifa_Zhang_NJP2016, Kamal_PRB_2021}
\be
\bm{M}_{\bm \Omega} = \frac{1}{\hbar} \sum_{\kb} \zeta \bm{\Omega}~.
\ee 
Here, $\zeta = - \int_{\e}^{\infty} d\eta (\eta -\mu)f(\eta)$ with $f(\eta)$ being the equilibrium Fermi distribution function and $\bm{\Omega}$ denotes the Berry curvature~\cite{Qian_Nu_ROMP_2010}.

We highlight that the contribution from the first term of Eq.~\eqref{Th_C_def} can be thought of as the equivalent of the semiclassical definition of the heat current ${\bm J}_s=\int [d{\bm k}] (\epsilon- \mu) g_{\bm r}$~\cite{SSP} where $g_{\bm r}$ is the temperature gradient induced non-equilibrium distribution function. The second term can be thought as the correction to the local current induced by the finite size of the wave packet. Moreover, the resultant contribution of the last two terms of Eq.~\eqref{Th_C_def} can be considered as the equivalent to the heat magnetization current density ($-{ \bm \nabla} \times {\bm M_Q}$)~\cite{Qian_Nu_PRB_2020}.
As a consistency check, using the definition in Eq.~\eqref{Th_C_def}, we have calculated the expression of the linear anomalous thermal Hall current (see Appendix~\ref{Appendix_LTC}) known in the semiclassical theory. 

To calculate the second-order heat current, we generalize Eq.~\eqref{Th_C_def} to include higher-order corrections in the density matrix due to temperature gradient. We define the $2^{\rm{nd}}$-order thermal current,  $\bm{J}^{(2)}$ in response to the temperature gradient as
\bea\label{Th_C_def_SC}
\bm{J}^{(2)}  &=& {\rm Tr }\Big[\frac{1}{2}\{\m{H}_0,\bm{v} \}\rho_{\rm D}^{(2)}\Big] + {\rm Tr }\Big[\frac{1}{2}\{\m{H}_0,\bm{v} \}\rho_{\rm O}^{(2)}\Big] \nn \\ 
&& +{\rm Tr}\Big[(\bm{E}_T \times \bm{m}_N) \m{H}_0 \rho^{(1)}_{\rm D}\Big]~.
\eea%
{Here, the first two terms arise from the trace of the energy current operator with the second-order density matrix. 
The third term in Eq.~\eqref{Th_C_def_SC} is the orbital magnetic moment energy current, which has to be subtracted from the local thermal current to obtain the physical transport thermal current. This term has $\rho^{(1)} \propto \nabla T$, as the local energy magnetization is itself proportional to a temperature gradient, making the net contribution $\propto (\nabla T)^2$. Finally, the Berry curvature energy magnetization term [the fourth term in Eq.~\eqref{Th_C_def}] is always linear in the temperature gradient. It does not contribute to the second-order thermal current calculation. 
}
%

\subsection{Solution of the density matrix in the presence of  temperature gradient}\label{den_Sol}

We can solve the density matrix equation given in Eq.~\eqref{QKL_eqn} perturbatively by expanding $\rho$ in  powers of the temperature gradient, 
$\rho= \rho^{(0)}+ \rho^{(1)} + \rho^{(2)} + \cdots $,
{where the $N$-th order correction to the density matrix (or the $N$-th order density matrix), $\rho^{(N)} \propto \vert ({\bm \nabla} T)^N \vert$. Note that this includes the cross terms of the temperature gradient such as $\rho^{(2)} \propto \partial_x T~ \partial_y T$.}
Here, the equilibrium density matrix is given by $\rho^{(0)} = \sum_n \vert u^n_{\kb}\rangle \langle u^n_{\kb} \vert f_{n}$, where $f_{n} \equiv f(\e_{n}) = \left(1+ e^{(\e_n-\mu)/k_BT} \right)^{-1}$ is the Fermi-Dirac distribution function and $k_B$ is the Boltzmann constant.
In the band basis, the $N$-th order density matrix can be calculated using the following equation (for more details of this equation, we refer the readers to appendix~\ref{app_adiabatic}),
{
\be \label{eqn:den3}
\frac{\partial \rho_{np}^{(N)}}{\partial t} +\frac{i}{\hbar}\left[ \m{H}_0,\rho^{(N)}\right]_{np}+ \frac{\rho^{(N)}_{np}}{\tau/N} = \left[ D_T (\rho^{(N-1)})\right]_{np}.  
\ee 
Here, the subscript `$np$' denotes the matrix element of a matrix 
sandwiched between the 
$n$-th and $p$-th energy eigenstates. For example, if ${\m O}$ is an arbitrary matrix, then ${\m O}_{np} \equiv \bra{u^n_{\kb}}{\m O} \ket{u^p_{\kb}}$.}
The first term in the above equation can be discarded in a steady state. Hereafter, we consider the steady state to calculate the density matrix. To calculate the first-order density matrix, we have computed the commutator relation $[\m{H}_0,\rho^{(1)}]_{np}= (\e_n -\e_p)\rho^{(1)}_{np} $ and the thermal driving term to be
\be
[ D_T (\rho^{(0)})]_{np} =
-\frac{1}{2\hbar}\bm{E}_T\cdot [ (\tilde{\e}_n 
+ \tilde{\e}_p){\bm \nabla}_{\kb}f_{n}\delta_{np} 
 + ~ 2i {\bm {\m{R}}}_{np} \xi_{np}]~.
\ee
Here, we have defined  $\xi_{np} = \tilde{\e}_n f_{n}- \tilde{\e}_p f_{p}$ to make the above expression simple.
Using these in Eq.~\eqref{eqn:den3}, we obtain the first order density matrix, 
\be \label{eqn:fod}
\rho^{(1)}_{np} =-  \left[\frac{\tau}{\hbar} \delta_{np} \tilde {\e}_n  \partial_c f_{n} +  {\m{R}}_{np}^{c}g^{np}_1\left(\tilde \e_nf_{n}- \tilde \e_p f_{p} \right)\right]E^c_T. 
\ee
%
For brevity, we have defined $\partial_{k_c} \equiv \partial_{c}$, and the Einstein summation convention for repeated spatial indices is used.
The first term of the density matrix corresponds to the diagonal element of the density matrix owing to the Dirac delta function ($\delta_{np}$), and it vanishes for insulators. The second term corresponds to the off-diagonal element, representing the inter-band coherence introduced by the temperature gradient. In Eq.~\eqref{eqn:fod}, we have defined $g^{np}_1 \equiv (\e_{np} -i\hbar/\tau)^{-1}$ with $\e_{np} \equiv (\e_{n}-\e_{p})$ being the interband energy gap at a given ${\bm k}$.  To keep track of the diagonal and the off-diagonal parts of the density matrix, we will use the notation $\rho^{\rm d}$, and ${\rho^{\rm o}}$ to denote the diagonal and the off-diagonal elements of the first order density matrix, respectively. 

Now, we calculate the second-order density matrix. Following the notation of the diagonal and the off-diagonal elements of the density matrix, 
we can express the second-order density matrix in four parts: two terms for the diagonal sector - $\rho^{\rm{dd}}, \rho^{\rm{do}}$, and two terms for the off-diagonal sector -  $\rho^{\rm{od}}, \rho^{\rm{oo}}$. Here, the first superscript denotes the diagonal or off-diagonal element of the second-order density matrix, while the second superscript represents the diagonal or off-diagonal element of the first-order density matrix involved.
For instances, $\rho^{\rm{dd}}$ denotes the diagonal part of $\rho^{(2)}$ arising from the diagonal part of $\rho^{(1)}$. Similarly, $\rho^{\rm{oo}}$ describes the off-diagonal part of the $\rho^{(2)}$ arising from the off-diagonal part of the $\rho^{(1)}$ and so on. 
Keeping the detailed calculation of the second-order density matrix in the appendix~\ref{s1}, we write the 
%
'dd' component of $\rho^{(2)}$ as 
\bea\label{rho_dd} 
\rho^{\rm{dd}}_{nn} =  \frac{\tau^2}{2\hbar^2} \left[ \hbar \tilde\e_n v^n_{b} \partial_c f_n  +  \tilde\e^2_n \partial_b \partial_c f_n \right] E^b_T E^c_T~,
\eea 
where, $v^n_b  = \hbar^{-1} \pdv{\e_n}{k_b}$ is the group velocity of the Bloch electrons in the $n$-th band along the spatial direction $b = x,y,z$. A similar expression of the diagonal part of the density matrix has been derived earlier using the semiclassical approach in Ref.~[\onlinecite{Xiao_PRB_2019}], where they calculated the second-order correction to the non-equilibrium distribution function due to the temperature gradient.
This density matrix component depends upon the Fermi function's derivative.  Therefore, this term governs the Fermi surface effect. 
%
The `do' component of $\rho^{(2)}$ is calculated to be, 
\bea 
\rho^{\rm{do}}_{nn} =&& \frac{i\tau}{4\hbar^2}\sum_{p}^{p \ne n} (\tilde{\e}_n + \tilde{\e}_p) \left(g^{np}_1\m{R}_{np}^{c}\m{R}_{pn}^{b} + g^{pn}_1\m{R}_{np}^{b}\m{R}_{pn}^{c}\right) \nn \\
&&\times~ 
\xi_{np} E^b_T E^c_T~.
\eea 
This part of the density matrix depends on the Fermi function and contributes as a Fermi sea effect. Continuing in chronological order, we calculate the `od' part of the second-order density matrix to be 
\bea 
\rho^{\rm{od}}_{np} &&= \frac{\tau}{\hbar}g^{np}_2 \m{R}_{np}^{b}\left(\tilde{\e}_n ^2 \partial_{c}f_{n}-\tilde{\e}_p^2 \partial_{c}f_{p} \right)  E^b_T E^c_T~,
\eea 
with $g^{np}_2 \equiv (\e_{np} -2i\hbar/\tau)^{-1}$. The dependency of this part of the density matrix on the derivative of the Fermi function implies the Fermi surface effects. Likewise, the `oo' part of $\rho^{(2)}$ is evaluated to be 
\bea \label{eq-11}
\rho^{\rm{oo}}_{np}  &=& \frac{-i}{2}g^{np}_2 (\tilde{\e}_n + \tilde{\e}_p ) \m{D}^{b}_{np}\left(g^{np}_1\m{R}_{np}^{c}\xi_{np}\right) E^b_T E^c_T \nn \\ 
&+&  \frac{1}{ 2}g_2^{np} \sum_{q\neq n \neq p} \left[ g^{nq}_1\m{R}^{c}_{nq}\m{R}^{b}_{qp}(\tilde{\e}_n +\tilde{\e}_q )\xi_{nq}\right. \nn \\ 
&-& \left. g^{qp}_1\m{R}^{b}_{nq}\m{R}^{c}_{qp}(\tilde{\e}_q +  \tilde{\e}_p )\xi_{qp}\right]E^b_T E^c_T~.
\eea 
Here, we have used $ \m{D}^b_{np} = \partial_b - i (\m{R}^b_{nn}-\m{R}^b_{pp} ) $. In Eq.~\eqref{eq-11},  the second term contributes only to systems having three or more bands. This completes our derivation of the second-order density matrix for a thermal perturbation to a multi-band system. 

\subsection{Calculation of the nonlinear thermal current}
In this section, we calculate the general expression of the NL thermal current as a second-order response in the temperature gradient. In Eq.~\eqref{Th_C_def_SC}, we showed that the thermal current depends on the diagonal and off-diagonal parts of the second-order density matrix along with the orbital magnetic moment contribution. To track the origin of different thermal current contributions, we split the total second-order thermal current as ${\bm J} = {\bm J}^{\rm dd} + {\bm J}^{\rm do} + {\bm J}^{\rm od} + {\bm J}^{\rm oo} + {\bm J}^{\rm OMM} $. Here, the different superscripts in the first four terms indicate the corresponding sectors of the second-order density matrix used for calculating the thermal current contribution. The superscript `OMM' convey the orbital magnetic moment-induced thermal current. For example, ${\bm J}^{\rm dd}$ means the thermal current originated from the second-order density matrix component $\rho^{\rm dd}$ and so on.  
Now in band basis form, we write the first term of Eq.~\eqref{Th_C_def_SC} along an arbitrary direction $a$ as 
\bea\label{firt_current}  
{\rm Tr}\left[ \frac{1}{2}\{ {\m H}_0, {\bm v}\} \rho^{(2)}_{\rm D}\right]_a = \sum_{n,\kb } \tilde\e_n  v^{n}_a \rho^{(2)}_{{\rm D},nn}~.
\eea 
For shorthand notation, we have defined $\sum_{\kb} \equiv \int [d\kb]$ with $[d\kb] = d^d\kb/(2\pi)^d$ being the integration measure for the $d$-dimensional system. In the previous section~\ref{den_Sol}, we have written the diagonal part of the second-order density matrix ($\rho^{(2)}_{\rm D}$) as a sum of two sectors $\rho^{\rm dd}$ and $\rho^{\rm do}$. So, $\rho^{(2)}_{\rm D}$ gives us two thermal current contributions through Eq.~\eqref{firt_current}.  The thermal current corresponding to $\rho^{\rm dd}$ is calculated to be
\be \label{NLD_therm}
J^{\rm dd}_{a} =  \frac{\tau^2}{2\hbar^2 } \sum_{n,{\bm k}}  ( \hbar \tilde\e_n v_b^n \partial_c f_n  +  \tilde\e^2_n \partial_{b}\partial_c f_n)  \tilde\e_n v^n_{a} E^b_T E^c_T~.  
\ee
A similar expression of thermal current can be calculated through the semiclassical Boltzmann framework, as shown in Ref.~[\onlinecite{Xiao_PRB_2019}]. Note that this current is quadratic in scattering time ($ \propto \tau^2$), it arises only due to the group velocity of the Bloch electrons, and it does not depend on any band geometric quantities. Therefore, this current is extrinsic, and it can be identified as the conventional \textit{NL thermal Drude current}. To the best of our knowledge, this is the first derivation of the NLD thermal current using the quantum kinetic theory. Further, we calculate the thermal current stemming from the $\rho^{\rm do}$ to be 
\bea\label{J_do}
J^{\rm{do}}_{a} &=& \frac{i\tau}{4\hbar } \sum_{n,p, {\bm k}}^{p\ne n} g^{np}_1 \m{R}^b_{pn}\m{R}^c_{np} (\tilde\e_n + \tilde\e_p) (\tilde{\e}_n f_{n}- \tilde{\e}_p f_{p}) \nn \\ 
&&\times \left(\tilde\e_n v^n_{a}- \tilde\e_p v^p_{a} \right) E^b_T E^c_T~.
\eea
The band geometric quantities are explicit in the above expression and unlike the conventional Fermi surface effect, this current is a Fermi sea effect. We  note that the structural form of this current is similar to the injection current known in photogalvanics~\cite{Atanasov_1996_PRL, Aversa_1995_PRB, Sipe_2000_PRB,Fregoso_2019_PRB, Sturman_2021_Book},
and it arises from energy injection $\left(\tilde\e_n v^n_{a}- \tilde\e_p v^p_{a} \right)$ across bands. 
Therefore, we will call it the \textit{thermal injection current}.

Similarly, the second term of Eq.~\eqref{Th_C_def_SC} is written in the band-reduced form as 
\be\label{second_current} 
{\rm Tr}\left[ \frac{1}{2}\{ {\m H}_0, {\bm v}\} \rho^{(2)}_{\rm O}\right]_a =  \frac{1}{2} \sum_{n,p,\kb}({\tilde \e}_n + {\tilde \e}_p)v^{pn}_a \rho^{(2)}_{{\rm O},np}~.
\ee 
Here, $v^{pn}_a$ denotes the components of the velocity operator defined as $i \hbar {\bm v}= [{\bm r}, {\mathcal H}_0]$. This can be obtained using the covariant derivative of the Hamiltonian ($ \bm{v}^{pn} = \hbar^{-1} [\m{D}_{\kb}(\m{H}_0)]_{pn}$) with the explicit form  
\be\label{velocity} 
v_a^{pn} = v_{a}^{n}\delta_{pn}+i\omega_{pn}\m{R}_{pn}^{a}~.
\ee
Here, the first term is the diagonal element of the velocity operator, which is equal to the gauge-invariant band velocity, while the second term is the off-diagonal component which consists of the band-resolved Berry connection and $\omega_{pn} = \e_{pn}/\hbar$ being the interband transition frequency. In principle, Equation~\eqref{second_current} gives us two more contributions of the thermal current when we put the different off-diagonal sectors ($\rho^{\rm od}$ and $\rho^{\rm oo}$) of the second-order density matrix. Using the form of $\rho^{\rm od}$, we calculate its corresponding thermal current to be
\bea\label{J_od} 
 J^{\rm{od}}_{a} & = & \frac{i\tau}{2\hbar^2} \sum_{n,p, \kb}^{p \ne n} \e_{pn} g^{np}_2 ( \tilde\e_n + \tilde\e_p) \m{R}^a_{pn}\m{R}^b_{np} \nn \\  
&& \times \left(\tilde\e_n^2 \partial_{c} f_{n} - \tilde\e_p^2 \partial_{c} f_{p} \right)  E^b_T E^c_T ~.
\eea 
%
Note that it is a Fermi surface current, and in the low-temperature limit, only states near the Fermi surface contribute to this current.
Finally, for the nonlinear thermal current stemming from $\rho^{\rm{oo}}$, explicit calculations yield  
\begin{widetext}
\bea \nn\label{J_oo} 
J^{\rm{oo}}_{a} &=&  -\frac{1}{4\hbar } \sum_{n,p,\kb}^{p \ne n}  \e_{np} g^{np}_2 ( \tilde\e_n + \tilde\e_p )\m{R}^a_{pn}\bigg[ ( \tilde\e_n + \tilde\e_p ) \m{D}^b_{np}( g^{np}_1 \m{R}^c_{np} \xi_{np})  
\\
& +&  i \sum_{q\neq n \neq p} \left( g^{nq}_1 \m{R}^{c}_{nq}\m{R}^{b}_{qp}(\tilde{\e}_n +\tilde{\e}_q )\xi_{nq} - 
g^{qp}_1 \m{R}^{b}_{nq}\m{R}^{c}_{qp}(\tilde{\e}_q +  \tilde{\e}_p )\xi_{qp} \right)\bigg]E^b_T E^c_T~.
\eea  
\end{widetext}
This completes the calculation of all the contributions to the NL thermal current arising from the first two terms in Eq.~\eqref{Th_C_def_SC}. 
So we are left with the contribution arising due to the orbital magnetic moment. After performing an explicit calculation of this term, we have  (See Appendix~\ref{Appendix_NLTC} for details)
\be\label{OMM_current}  
J_{a}^{\rm{OMM}}  = \frac{\tau }{2 \hbar^2} \sum_{n,p,\kb}^{p \ne n}  \tilde\e_n^2 (\tilde\e_n - \tilde\e_p) \Omega^{ab}_{np} \partial_c f_n E^b_T E^c_T~.
\ee  
Here, the quantity $\Omega^{ab}_{np}$ is a band geometric quantity commonly known as the band resolved Berry curvature. It is the imaginary part of the product of the band-resolved Berry connection and is defined as $\Omega^{ab}_{np} = i[\m{R}^a_{np}\m{R}^b_{pn} - \m{R}^b_{np}\m{R}^a_{pn}] $. We emphasize that this is antisymmetric under exchange of both spatial or band indices $i.e. ~ \Omega^{ab}_{np} = - \Omega^{ba}_{np} ~ {\rm{and}} ~ \Omega^{ab}_{np} = -\Omega^{ab}_{pn}$~\cite{xiao_RMP2010, Watanbe_PRX_2021}. 

Now, we show that our results are consistent with the general symmetry requirements. Since we are calculating the second-order NL thermal response, we should first consider the space inversion symmetry (or parity symmetry). A general symmetry analysis of $J_a=-\kappa_{abc}\nabla_b T \nabla_c T$ shows that the second-order responses ($\kappa_{abc}$) should vanish in the presence of parity symmetry. To our satisfaction, we find that the derived expressions are consistent with this symmetry restriction. In the presence of parity symmetry, the energy dispersion, Berry connection, Berry curvature, and the group velocity satisfy the relations $\e(\kb) = \e(-\kb),~ \m{R}^a_{np}(\kb) = -\m{R}^a_{np}(-\kb),~\Omega^{ab}_{np}(\kb) = \Omega^{ab}_{np}(-\kb)~ {\rm and }~ v^n_a (\kb) = -v^n_a(-\kb) $, respectively. Using these relations in the calculated thermal current, we find that all the components of the NL thermal current vanish over Brillouin zone (BZ) integration under $\m{P}$-symmetry.

%
\begin{table*}[t]
    \centering
    \caption{Second order nonlinear thermal current for a two-band system. Each of the contributions can be expressed as $J_a = \hbar^{-2} \sum_{n,p} \int [d\kb] E^b_T E^c_T  \times A \times B$, where $A$ and $B$ are listed below. Specifically, the impurity-dependent term, $B = B_{\rm int} + B_{\rm ext}$, has been expressed as a sum of an intrinsic (impurity independent) and extrinsic (impurity dependent) term. The last two columns indicate which of these contributions survives in the presence of either time-reversal symmetry or the presence of parity-time reversal symmetry. Here $(n,~p)$  denote the band indices, $(a, ~b, ~c)$ denote the cartesian indices, $\tilde \varepsilon = \varepsilon - \mu$ and $\cal R$ denotes the Berry connection. Furthermore, the interband transition frequency is defined as $\hbar \omega_{np} = (\tilde\e_n - \tilde\e_p) \equiv \e_{np}$. 
    } 
    \label{current_table}
    \begin{tabular}{ccc cc}
    \hline \hline 
    \rule{0pt}{3ex}
    Current & $A$ &$ B  =  B_{\rm int} +    B_{\rm  ext}$ & $\m{T}$-symmetry & $\m{PT}$-symmetry \\[1ex]
    \hline \hline 
    \rule{0pt}{3ex}
      $J^{\rm{dd}}_a$   & $\frac{1}{2} (\hbar \tilde\e_n v^n_b \partial_c f_n  +  \tilde\e^2_n \partial_{b}\partial_c f_n)  \tilde\e_n v^n_{a} \delta_{np} \times $ & [ 0 $+ \tau^2   $] & 0 & $\neq 0 $  \\ [2ex]
      $J^{\rm{do}}_a$ &  $-\frac{\tilde\e_n (\tilde\e_n+ \tilde\e_p)}{2\omega_{np}^{2} } \left(\tilde{\e}_n v^n_{a}-\tilde{\e}_p v^p_{a}\right) f_n \times $ & [$\m{G}^{bc}_{np} + 0] $  &  0 & $\neq 0 $ \\[2ex]
      $J^{\rm{od}}_a + J^{\rm OMM}_a$ &  $ \tilde\e_n^2     \partial_{c} f_{n} \times  $ &  $[ \frac{2(\tilde\e_n+ \tilde\e_p)}{\omega_{np}} \m{G}^{ab}_{np} +  \tau \tilde\e_n  \Omega^{ab}_{np}] $  &  $\neq 0$  &  $\neq 0 $  \\ [3ex]
      $J^{\rm{oo}}_a$ & $\frac{1}{4}  ( \tilde\e_n + \tilde\e_p )^2 \m{R}^a_{pn} \times $ & $[-\m{D}^b_{np}\left( \frac{\m{R}^c_{np} \xi_{np}}{\omega_{np}}\right) + 0]$ & $0$ & $\neq 0$ \\ [3ex]
      %
    \hline 
    \end{tabular}
\end{table*}

\subsection{Intrinsic and extrinsic thermal current}
In this section, we will separate the different scattering time dependence of the thermal current in the form ${\bm J}= {\bm J}( \propto \tau^{0}) + {\bm J}(\propto \tau^{1})+ {\bm J} (\propto \tau^{2})$. Here, ${\bm J}( \propto \tau^{0})$ is the intrinsic part which represents the dissipationless thermal current, and the other two terms represent extrinsic contributions. This type of classification has certain advantages. First of all, in the experiments, the different scattering time dependents can be separated using appropriate scaling laws~\cite{du_NatCom2019_disorder}. This, in turn, helps us to understand and separate the physical mechanism behind the dominant contribution to the thermal current. Moreover, this kind of classification helps us to identify dissipationless NL thermal currents, which originate from quantum coherence effects and are very fascinating. Finally, it is much easier to compare results from quantum kinetics with the semiclassical thermal transport, where currents are calculated in various orders of the scattering time. 

We note that the scattering time dependence of the contributions ${\bm J}^{\rm dd}$ and ${\bm J}^{\rm OMM}$ are $\tau^2$ and $\tau$, respectively. However, the scattering time dependence of the remaining current is not trivial and is determined by the factor $\tau g^{np}_1$ for ${\bm J}^{\rm do}$, $\tau g^{np}_2$ for ${\bm J}^{\rm od}$, and $g^{np}_1 g^{np}_2 $ for ${\bm J}^{\rm oo}$. To 
extract the scattering time dependence of the terms from these factors, we use the following identities:
\be\label{Identity}
\frac{\tau}{i\hbar} g^{np}_N = \frac{1}{\e^2_{np}}\left( N + \eta^{np}_{\tau,N}\right) \ \ \mbox{and} \ \ g^{np}_N  = \frac{1}{\e_{np}}(1 + \tilde{\eta}^{np}_{\tau,N}) .
\ee
Here, we have used the generalized expression of $g^{np}_1$ and $g^{np}_2$ in the form of $g^{np}_{N} \equiv (\e_{np} - N\frac{i\hbar}{\tau})^{-1}$. 
In the above equation, $\eta^{np}_{\tau,N}$ and $\tilde{\eta}^{np}_{\tau,N}$ are a dimensionless function of $\tau\omega_{np}$ and their explicit dependence is given as
\be
\bal
& \eta^{np}_{\tau,N}=  -i \tau \omega_{np} \left( \frac{ 1 -i N^3 \frac{1}{\tau^3 \omega_{np}^3} }{1 + N^2 \frac{1}{\tau^2 \omega^2}} \right)~,  \\
& \tilde\eta^{np}_{\tau,N} = N\frac{i}{\tau \omega_{np}} \left(\frac{1 + N \frac{i}{\tau \omega_{np}}}{1 + N^2 \frac{1}{\tau^2 \omega^2_{np}}}\right)~.
\eal
\ee
Using these identities, we have separated the intrinsic and extrinsic parts of the NL current in Table.~\ref{current_table}. Further, to obtain the thermal current of various orders of $\tau$ dependencies, we consider the dilute impurity limit (DIL) where we consider the scattering time ($\tau$) to be much greater than the inverse of the interband transition frequency $\omega_{np}$, i.e., $\frac{1}{\tau \omega_{np}} \ll 1 $. In this limit, ignoring the higher order terms in $(\tau \omega_{np})^{-1}$, we can write $\tau g^{np}_N \approx \frac{i\hbar N}{\e_{np}^2} + \frac{\tau}{\e_{np}}$ and $g^{np}_N \approx \frac{1}{\e_{np}}$. For the detailed calculation of these identities, we refer readers to Appendix~\ref{the identities}. Using these identities and within DIL approximation, we can separate the NL thermal currents depending on different powers of $\tau$ as,
\be\label{TC_tau_dependence}
\begin{aligned}
J_a(\propto \tau^0) &= J^{\rm do}_{a,\rm int} + J^{\rm od}_{a,\rm int} + J^{\rm oo}_{a,\rm int}~, \\ 
J_a(\propto \tau^1) &= J^{\rm do}_{a,\rm ext} + J^{\rm od}_{a,\rm ext}+ J^{\rm OMM}_{a}~, \\ 
J_a(\propto \tau^2) &= J^{\rm dd}_{a}~.
\end{aligned}
\ee
Here, the subscript ${\rm int}~ ({\rm ext})$ conveys the intrinsic (extrinsic) part of the thermal current. {Note that $J_a(\propto \tau^0)$ is the only nonlinear current which survives in the ballistic limit.}
As expected, the $J^{\rm dd}_{a}$ current gives rise to the quadratic scattering time dependence, and $J^{\rm OMM}_{a}$ contributes to the linear scattering time-dependent current. From the dilute limit expansion of the scattering time-dependent factor $\tau g^{np}_1$, one expects an intrinsic and a linear scattering time-dependent term from the ${\bm J}^{\rm do}$ component of the thermal current. Surprisingly, we find that the extrinsic part of the ${\bm J}^{\rm do}$ component vanishes identically, i.e. ${\bm J}^{\rm do}_{\rm ext}=0$. Hence, we are left with the intrinsic part given by, 
\be
J^{\rm{do}}_{a,\rm{int}} = - \frac{1}{ 2 } \sum_{n,p,\kb}^{p \ne n} \frac{\m{G}^{bc}_{np}}{\e_{np}^2} \tilde\e_n (\tilde\e_n + \tilde\e_p) \left(\tilde{\e}_n v^n_{a}-\tilde{\e}_p v^p_{a}\right) f_n E^b_T E^c_T~,
\ee 
Here, $\m{G}^{bc}_{np}$ is band resolved quantum metric tensor and is defined as $\m{G}^{bc}_{np} = \frac{1}{2}[\m{R}^b_{np}\m{R}^c_{pn}+\m{R}^c_{np}\m{R}^b_{pn}]$. Note that it is symmetric under the exchange of both spatial ($b,~c$) and band ($n,~p$) indices, i.e. $\m{G}^{bc}_{np} = \m{G}^{cb}_{np} ~ {\rm{and}} ~ \m{G}^{bc}_{np} = \m{G}^{bc}_{pn}$~\cite{Provost_CIMP_1980,M_Berry_1989,Watanbe_PRX_2021}. 
Continuing with the same approach, we obtain non-zero intrinsic and linear scattering time-dependent extrinsic contributions from the $J^{\rm od}_a$ thermal current component. The intrinsic part is determined by the band-resolved quantum metric and calculated to be
\be \label{J_od_int}
J^{\rm{od}}_{a,\rm{int}} = \frac{2}{\hbar } \sum_{n,p,\kb}^{p \ne n} \frac{\m{G}^{ab}_{np}}{\e_{np}} \tilde\e_n^2 (\tilde\e_n+ \tilde\e_p)   \partial_{c} f_{n}  E^b_T E^c_T~.
\ee
The extrinsic contribution is determined by the band-resolved Berry curvature and given by, 
\be\label{J_od_extrinsic} 
J^{\rm{od}}_{a,\rm{ext}} = \frac{\tau}{2 \hbar^2} \sum_{n,p,\kb} \tilde\e^2_n (\tilde\e_n+ \tilde\e_p) \Omega^{ab}_{np} \partial_c f_n E^b_T E^c_T~.
\ee 
Finally, we calculate the intrinsic contribution of the $J^{\rm oo}_a$. After performing a straightforward algebra, we obtain the simplified expression of the intrinsic thermal current originating from $J^{\rm oo}_a$ as  
\be\label{J_OO_intrinsic} 
\bal
J^{\rm oo}_{a,\rm{int}} =& \frac{1}{4\hbar} \sum_{n,p,\kb} ( \tilde\e_n + \tilde\e_p )\frac{\tilde\e_n f_n}{\e_{np}} \bigg[ ( \tilde\e_n + \tilde\e_p ) \partial_a {\m G}^{bc}_{np} \\ 
& + 4 {\m G}^{ac}_{np} \partial_b ( \tilde\e_n + \tilde\e_p )\bigg]E^b_T E^c_T~.
\eal
\ee 
We refer the reader to appendix~\ref{appendix_for_joo}, where we have presented a detailed calculation of this intrinsic current. This summarizes all the components of        Eq.~\eqref{TC_tau_dependence}. The details of calculation for the intrinsic and extrinsic parts of the different NL thermal current components are presented in Appendix~\ref{ext_and_int}.

We find that the linear scattering time-dependent currents given in Eq.~\eqref{OMM_current} and Eq.~\eqref{J_od_extrinsic} can be combined to write in a compact form. The resultant expression is given by 
\be\label{Th_anomalous_cur} 
J_a(\propto \tau^1) = \frac{\tau}{\hbar^2} \sum_{n,p,\kb} \tilde\e^3_n \Omega^{ab}_{np} \partial_c f_n E^b_T E^c_T~.
\ee 
Remarkably, this expression is known as the  \textit{NL anomalous (NLA) thermal current} in literature. This Hall contribution to the NL thermal current has been obtained earlier using the semiclassical Boltzmann transport framework in Refs.~[\onlinecite{Nandy20, zhou_PRB2022,chakraborty_2DM2022_non}] and viewed as the thermal analog of the Berry curvature dipole. Here, we have obtained this using the quantum kinetic theory framework. 

\begin{table}[h]
    \centering
    \begin{tabular}{c c c c}
    \hline \hline
         Quantities & ${\m P}$ & ${\m T}$ & ${\m PT}$  \\ [1ex]
         \hline \hline
         $\e_n(-\kb)$& $\e_n(\kb)$ & $\e_n(\kb)$ & --- \\ [0.5ex]
         ${\bm v}_n(-\kb)$ & $ - {\bm v}_n(\kb)$ & $-{\bm v}_n(\kb)$ &--- \\ [0.5ex]
         ${\m R}^{a}_{np}(-\kb)$ ~~~& ~~~$-{\m R}^{a}_{np}(\kb)$~~~ &~~~ ${\m R}^{a}_{pn}(\kb)$~~~&~~~ $-{\m R}^{a}_{pn}(-\kb)$ \\ [0.5ex]
         $ \Omega^{ab}_{np}(-\kb)$ & $\Omega^{ab}_{np}(\kb)$ & $-\Omega^{ab}_{np}(\kb)$ & $-\Omega^{ab}_{np}(-\kb)$ \\ [0.5ex]
         ${\m G}^{ab}_{np}(-\kb)$ &${\m G}^{ab}_{np}(\kb)$ &${\m G}^{ab}_{np}(\kb)$ & ${\m G}^{ab}_{np}(-\kb)$ \\[0.5ex]
         \hline \hline 
    \end{tabular}
    \caption{Transformation of different physical quantities under various symmetries}
    \label{Symmetry_Table}
\end{table}

The separation of the heat current into the intrinsic and extrinsic parts allows us to do a time-reversal symmetry (TRS) analysis. A general symmetry analysis of $J_a = -\kappa_{abc}(\tau)\nabla_b T \nabla_c T$ implies that in systems with TRS, only the odd power of the $\tau$-dependent thermal currents survive. It can be understood as follows. Under time reversal, the thermal current $J_a$ and scattering time $\tau$ change their sign while $\nabla_b T \nabla_c T$ remains unaltered. Thus, in the presence of TRS, we have $\kappa_{abc}(-\tau) = -\kappa_{abc}(\tau)$. This condition implies that $\kappa_{abc}$ vanishes when it depends on the even powers of $\tau$. Therefore, we expect that the thermal current contributions $\propto \tau^0$ and $\propto \tau^2$ given in Eq.~\eqref{TC_tau_dependence} will vanish in the presence of TRS. To our satisfaction, using the relations given in Table~\ref{Symmetry_Table}, we find that only the thermal current contribution linear in the scattering time given in Eq.~\eqref{Th_anomalous_cur} is nonzero, while the other contributions vanish in time-reversal symmetric systems. Furthermore, we find that the parity-time reversal symmetry, when both the parity and time-reversal symmetries are individually broken, can have nontrivial consequences on the thermal current. From Table~\ref{Symmetry_Table}, it is evident that in presence of parity-time reversal symmetry, the Berry curvature vanishes at each point in the momentum space. Hence, the NL thermal currents depending on the Berry curvature vanish. So, we are left with the trivial $\propto \tau^2$  NL Drude current, along with an intrinsic scattering time-independent contribution to the thermal current.

\begin{figure}[t]
    \centering
    \includegraphics[width = \linewidth]{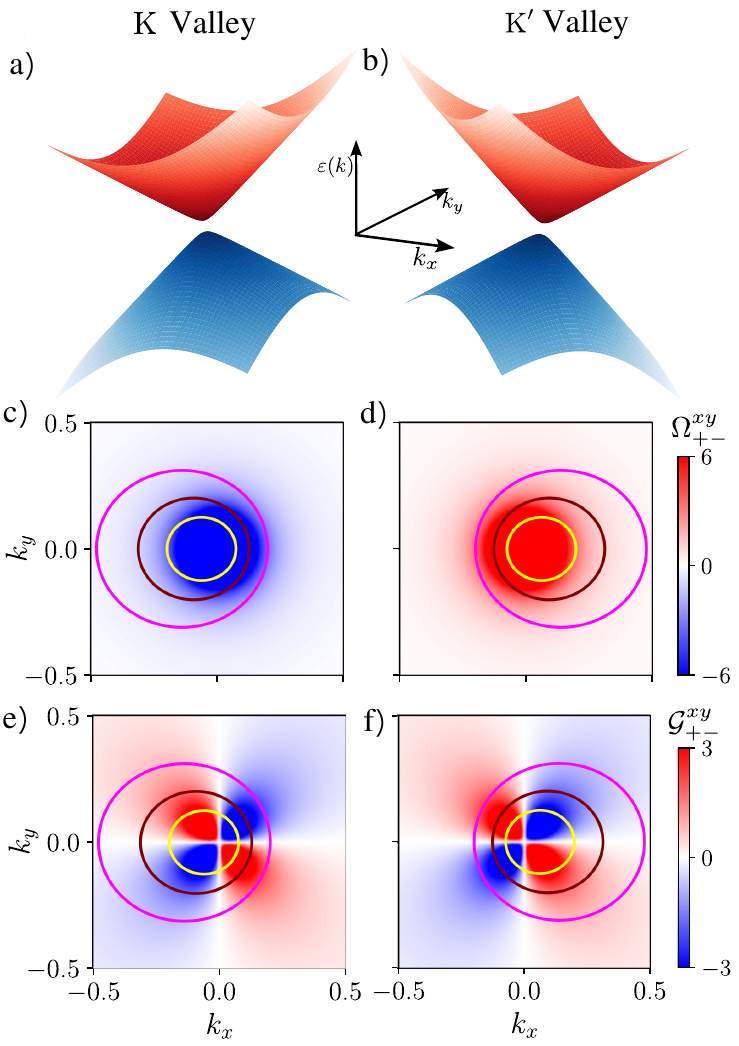}
    \caption{Energy dispersion of the tilted massive Dirac system specified by  Eq.~\eqref{Hamiltonian} is plotted against $k_x$ and $k_y$  in a) for the K valley and in b) for the K$'$ valley. c) and d) display the momentum space distribution of the band-resolved Berry curvature $\Omega^{xy}_{+-}$ for the two valleys, respectively. The momentum space distribution of the band-resolved quantum metric ${\m G}^{xy}_{+-}$ for the two valleys are shown in panels e) and f), respectively. The different energy contours in c)-f) are indicated by yellow ($\mu = 0.2$ eV), marron ($\mu = 0.3$ eV) and magenta ($\mu = 0.45$ eV) lines. Here, $k_x$ and $k_y$ are in the units of {\AA}$^{-1}$ and $\e (k)$ in eV. For these plots, we have used $\hbar v_F = 1.55$ eV{\AA}, $v_t = 0.4v_F$ and $\Delta = 0.1$ eV.}
    \label{fig:fig2}
\end{figure}

\section{Model specific calculation}
\label{sec:SecIII}

To study the NL thermal currents, we consider the massive tilted Dirac Hamiltonian \cite{Sadhukhan_2017, Politano_2018}. This model has been earlier used to explore the NL charge  transport~\cite{Nature2019, Huiying_Liu_PRB2022, Hangchao_Li_arxiv2020, Sodemann2015}. Its low-energy Hamiltonian is represented by~\cite{mojarro_PRB2021_optic, mojarro_PRB2022_hyper}
\bea \label{Hamiltonian}
\hat{\m{H}} = s  v_t k_x \sigma_0 + v_F(sk_x \sigma_x + k_y \sigma_y) + \Delta \sigma_z
\eea
Here, $\sigma_{x,y,z}$ are the Pauli matrices and $\sigma_0$ stands for a $2\times 2$ unit matrix, $v_F$ is the Fermi velocity and $v_t$ is the tilt velocity. The valley index $s = 1~(s = -1) $ represents the K (K$'$) valley. The mass term $\Delta$ opens a gap of $2\Delta$ between the conduction and valence bands and breaks the parity symmetry. 
%
%
We emphasize that for this Hamiltonian, the $y$-direction mirror symmetry that converts $k_y \rightarrow -k_y$ is preserved, while the $x$-direction mirror symmetry that converts $k_x \rightarrow -k_x$ is broken. As we will see later, mirror symmetry is crucial to determine the non-zero Berry curvature dipole in the system. The tilted Dirac fermions specified by the Hamiltonian in Eq.~\eqref{Hamiltonian} can be observed on the surface of topological crystalline insulators. In particular, topological crystalline insulators such as SnTe, Pb$_{1-x}$Sn$_x$Te, and Pb$_{1-x}$Sn$_x$Se can potentially host massive Dirac  fermions~\cite{Junwei_PRB2013}. The transition-metal dichalcogenides also host the Dirac cones~\cite{Bahramy_Nature2018} which can be tilted in presence of strain.

The energy dispersion for the Hamiltonian in Eq.~\eqref{Hamiltonian} is given by $\e_{\pm}(\kb) =s k_x v_t \pm \e_0$, where $\e_0 = \sqrt{\Delta^2 + v_F^2k^2}$ with $k^2 = k_x^2 + k^2_y $. The $+~(-)$ sign stands for the conduction (valence) band. The energy dispersion is shown in Fig.~\ref{fig:fig2}(a)-(b), where the left panel shows the K-valley and the right panel shows the K$'$-valley. Note that due to the valley index-dependent tilt, the Dirac cones are oppositely tilted. The anisotropic dispersion results in different band velocities: $v^{\pm}_{0,x} = s v_t \pm  v_F^2 k_x/\e_0$ along the $x$-direction, and  $v^{\pm}_{0,y} = \pm v_F^2  k_y/\e_0$ along the $y$-direction.
The band geometric quantities, such as Berry curvature and quantum metric, for this Hamiltonian, are given by,  
\begin{equation} \label{Geometric_quantities}
\begin{aligned}
\Omega^{xy}_{\pm,\mp} &= \mp \frac{s~\Delta v_F^2}{2 \e_0^3}~, \\ 
\m{G}^{xx}_{\pm,\mp} &= \frac{ v_F^2(\Delta^2 + k_y^2 v_F^2) }{4\e_0^4}~, \\
\m{G}^{xy}_{\pm,\mp} &= -\frac{k_x k_y v_F^4 }{4\e_0^4}~, \\
\m{G}^{yy}_{\pm,\mp} &=  \frac{ v_F^2(\Delta^2 + k_x^2 v_F
^2) }{4\e_0^4}~.
\end{aligned}
\end{equation}
There are a few important points that we can infer from these expressions. First of all, expressions of the band-resolved Berry curvature and quantum metric are independent of the tilt parameter $v_t$ and both of these quantities are highly concentrated near the band minima or maxima. We note that while the Berry curvature is valley contrasting, the quantum metric is the same in both valleys. Furthermore, the Berry curvature vanishes for gapless systems ($\Delta=0$) while the quantum metric is finite even in absence of a gap. Note that the expression of the Berry curvature is consistent with those in  Refs.~[\onlinecite{ZZ_Du_PRL2018, Toshihito_NJPS2020, Carmine_AQT2021}], and the various components of the quantum metric tensor are identical to those obtained in Ref.~[\onlinecite{Huiying_Liu_PRL2021, Huiying_Liu_PRB2022}]. 
We have shown the momentum space distribution of the band-resolved Berry curvature $\Omega^{xy}_{+-}$ for both the valleys in Fig.~\ref{fig:fig2}(c)-(d). In Fig.~\ref{fig:fig2}~(e)-(f), we have shown the momentum space distribution of the quantum metric tensor component ${\m G}^{xy}_{+-}$ for both the valleys. 

\begin{figure}[t!]
\centering
\includegraphics[width = \linewidth]{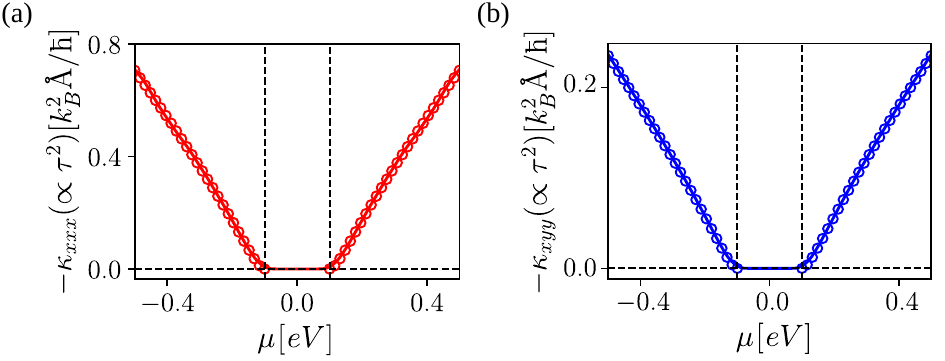}
\caption{Variation of the a) longitudinal and b) transverse nonlinear Drude conductivity with chemical potential. The solid line represents the numerical results, while the circles convey the analytical results calculated up to linear order in tilt velocity $v_t$. For numerical analysis, we have used $\hbar v_F = 1.0$ eV-{\AA}, $v_t = 0.1 v_F$, $\Delta = 0.1$ eV, and $\hbar/\tau = 0.16$ eV. In addition, we have considered the temperature to be $50$ K.}
\label{fig:fig3}
\end{figure}

We now use the expressions derived in Eq.~\eqref{Geometric_quantities} to calculate the different components of the current.
For a two-dimensional system, there can be six independent elements of NL thermal conductivity.  For simplicity, we assume an external temperature gradient along the $x$-direction and probe longitudinal and Hall-like thermal currents. The longitudinal current along the $x$-direction is determined by the conductivity $\kappa_{xxx}$, and the NL thermal current along the $y$-direction is determined by $\kappa_{yxx}$. For all the currents, we will focus on these two elements of NL thermal conductivity. We will start with the $\tau^2$ NLD. The NLD conductivity for a single Dirac node is calculated to be
\be
\kappa^{\rm NLD}_{xxx} = 3\kappa^{\rm NLD}_{yxx} =- sign(\mu) s \frac{\mu v_t \pi \tau^2}{8}(1-r^2 )^2 \Theta(\abs{\mu} - \abs{\Delta})~.
\ee
Here, we have defined $r = \Delta/\mu$ and $\Theta(\abs{\mu} - \abs{\Delta})$ as the Heaviside step function. We emphasize that for obtaining the analytical expressions, we have restricted ourselves to the linear order in the tilt velocity $v_t$. We note, unlike the linear Drude thermal conductivity, the NL Drude thermal conductivity is independent of temperature. Furthermore, the NL conductivity is valley index dependent which originates from linear tilt dependence. This implies that for Dirac nodes with opposite tilt, the total NL Drude contribution vanishes. This can be justified by the fact that in presence of time reversal symmetry which restricts the nodes to be oppositely tilted, the NL Drude conductivity is expected to vanish. In Fig.~\ref{fig:fig3}, we have shown the variation of the NLD conductivity with the chemical potential ($\mu$). Both the longitudinal and transverse conductivity increase with the chemical potential. The NLD conductivity keeps its sign preserved in both bands and reduces to zero when chemical potential reaches the band gap. Since the NLD conductivity is a Fermi surface effect so it is expected to vanish inside the gap.

\begin{figure*}[t!]
\centering
\includegraphics[width = \linewidth]{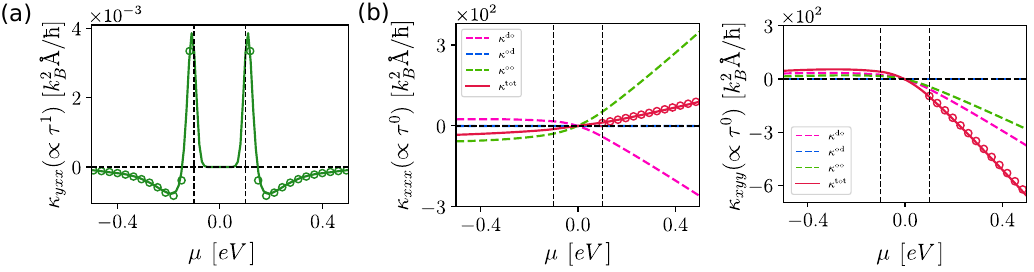}
\caption {(a) Variation of the nonlinear anomalous Hall conductivity with the chemical potential $\mu$ where solid line represents the numerically obtained result. In (b), we have shown the variation of the  longitudinal (left) and transverse (right) total nonlinear intrinsic conductivity with chemical potential ($\mu$). The total nonlinear intrinsic conductivity $\kappa^{\rm tot}$ is the sum of the intrinsic contributions arising from $\kappa^{\rm do}$, $\kappa^{\rm od}$ and $\kappa^{\rm oo}$. Therefore, in (b), the solid line represents the total intrinsic conductivity, while the dashed lines are the different thermal current contributions giving rise to intrinsic conductivity. Additionally, in this figure, the circles denote the analytical result calculated up to linear order in tilt velocity $v_t$. The model parameters, scattering time, and temperature considered here are the same as in Fig.~\ref{fig:fig3}. For the numerical and analytical calculation of the nonlinear intrinsic thermal current, we considered the cut-off energy of the valence band to be $2 \ {\rm eV}$.}
\label{fig:fig4}
\end{figure*}

Now, we turn our focus to the NL anomalous thermal current given in Eq.~\eqref{Th_anomalous_cur}. This linear $\tau$-dependent thermal current contribution has a nonzero value only in the direction perpendicular to the applied temperature gradient. The analytical expression for the $\kappa_{yxx}$ component of the anomalous conductivity, up to linear order in $v_t$, is given by 
\be 
\kappa^{\rm NLA}_{yxx} = -sign(\mu)\frac{7\pi^3 v_t k_B^2 T^2}{20 \mu^2} \left(  \frac{k_B^2}{\hbar} \right) r (1-2r^2) \Theta(\abs{\mu} - \abs{\Delta})~.
\ee 
This expression is independent of the valley index, and unlike the NLD conductivity, it depends on temperature quadratically. The chemical potential dependence of this conductivity is shown in Fig.~\ref{fig:fig4}\tc{blue}{(a)} with a good match between the analytical and numerical results, where we observe the following features. Firstly, the conductivity peaks near the conduction and the valence band edge and vanishes when we move away from the band edges. In addition to that, a change in the sign of the conductivity can be observed near $|\mu| = \sqrt{2} \Delta $. It attains a minimum value at $ |\mu | = \sqrt{6}\Delta$; thereafter, it becomes zero with increasing $\mu$. 


Finally, we calculate the analytical expressions of the NL intrinsic (NLI) thermal current given in Eq.~\eqref{J_od_int} and Eq.~\eqref{J_OO_intrinsic}. {For this, we consider the Fermi level to reside in the conduction band. Since the expressions of intrinsic conductivities consist of the Fermi function, therefore we have to consider the total contribution from the valence band in addition to the contribution from the conduction band. For the tilted Dirac model, the energy of the conduction band goes from $\Delta$ to $\infty$, while the valence band spans $-\Delta$ to $-\infty$. Having the chemical potential in the conduction band provides an energy cut-off for the conduction band electrons through the Fermi function. In this case, the valence band is completely filled, and the whole valence band contributes to the conductivity. In numerical simulations, we can not cover the whole valence band as it goes to infinity. To remedy this, we consider a cut-off in energy $-\Lambda$ for the valence band, with $\Lambda$ being a positive real quantity.} With this approach, we calculated the analytical result for the longitudinal and transverse intrinsic currents. The analytical form of the longitudinal intrinsic current  is 
%
\be\label{Int_analytical_xxx} 
\kappa^{\rm NLI}_{xxx} = \frac{s \mu v_t}{128 \pi  k_B^2 T^2} \left(  \frac{k_B^2}{\hbar} \right) \left[ 4 - 2r^2  -3 \lambda_2 - 4\lambda_1^2 \right]~, 
\ee 
where $\lambda_1 = \Delta/\Lambda$ and $\lambda_2 = \mu/\Lambda$. {Note that as $\Lambda \to \infty$, we have $\lambda_1 \to 0$ and $\lambda_2 \to 0$}. Similarly, the analytical expression of the transverse intrinsic current is 
%
\be\label{Int_analytical_yxx}
 \kappa^{\rm NLI}_{xyy} = \frac{-s \mu v_t}{128 \pi  k_B^2 T^2} \left(  \frac{k_B^2}{\hbar} \right) \left[ 20 - \frac{2}{3}r^2 - \frac{1}{2} r^4 - 25 \lambda_2 \right]~. 
%
\ee
These analytical expressions of the nonlinear intrinsic current are valid only when the chemical potential lies in the conduction band. Note that these expressions explicitly contain the valley index implying that for systems with time-reversal symmetry, the total intrinsic contribution from both of the nodes will vanish. 
We have shown the numerical results of the intrinsic conductivities in Fig.~\ref{fig:fig4}\tc{blue}{(b)} where we find a good match between the analytical and numerical results.  Note that the constant relaxation time approximation used by us is likely to work in the low-temperature regime, where the phonons and other thermal fluctuations are `frozen'.

{We expect the nonlinear thermal Hall  responses to dominate in systems preserving ${\cal T}$ symmetry, so that the linear Hall responses vanish in such systems. However, even in ${\cal T}$ broken systems, the linear and non-linear response signals can possibly be distinguished via lock-in frequency measurements. 
To get an order of magnitude estimate of the different current contributions for the model considered above, we set $\mu$ = 0.2 eV, $\tau = 10^{-12}$ s, and\cite{Guin2019}  $\partial_x T  = 1 K/mm $. With these parameters,  we find that in the linear response regime, we have the Drude heat  current $J^{\rm D}_x =-\kappa_{xx}^{\rm D} \partial_x T = -5.40 \times 10^{-9} \ {\rm W/mm}  $, and the linear anomalous Hall heat current, $J^{\rm AHE}_y = -  \kappa_{yx}^{\rm AHE} \partial_x T = -7.72 \times 10^{-12} \ {\rm W/mm} $. Similarly, for the non-linear response regime, we have the non-linear Drude heat current $J^{\rm NLD}_x = -\kappa_{xxx}^{\rm NLD} (\partial_x T)^2 = 2.08\times 10^{-15} \ {\rm W/mm} $, and the nonlinear Anomalous Hall heat current $J^{\rm NLA} = -\kappa_{yxx}^{\rm NLA} (\partial_x T)^2 = 3.32\times 10^{-20} \ {\rm W/mm} $. Similarly, we can calculate the nonlinear intrinsic Hall heat current to be $j^{\rm NLI} = - \kappa_{xyy} (\partial_y T)^2 = 4.05 \times 10^{-15} \ {\rm W/mm}$.}

\section{Discussions}
{For the sake of simplicity of analytical calculations, we have used the constant relaxation time approximation. However, we can easily include the momentum and energy dependence of the scattering time in our framework. For example, in Eq.~\eqref{firt_current} of our manuscript, the momentum dependent $\tau$ can be taken inside the Brillouin zone sum. The energy and momentum dependence of $\tau$ and the Brillouin zone sum can be evaluated numerically for any given system with a tight binding or continuum Hamiltonian. }

{We note that recently, there have been predictions of the nonlinear intrinsic electrical Hall conductivity. In Refs.~[\onlinecite{gao_PRL2014_field, nandy_PRB2019_symmetry, Huiying_Liu_PRL2021, wang_PRL2021_intrin}], an intrinsic nonlinear electrical Hall current resulting from the Berry connection polarizability (BCP) has been explored. More recently, it has been shown that additional new terms contribute to the nonlinear intrinsic current including a longitudinal non-Hall type nonlinear current [\onlinecite{shibalik_intrinsic_2022, gangsu_arX2022_intrinsic}], in addition to the BCP Hall current. Our paper studies the thermal version of this problem, and it explores both the intrinsic Hall and longitudinal thermal currents. }

{Another aspect about our work is that it is based on abelian Berry connection. This works as long as the bands are non-degenerate. In case of materials with composite ${\cal P} \cal {T}$ symmetry, all the bands are doubly degenerate. To be mathematically correct, for such cases our calculations should be generalized to include non-abelian Berry connection. However, our abelian Berry connection framework can still be used in a limiting sense for systems with the ${\cal P} \cal {T}$  being violated by a negligibly small parameter.} 

{The nonlinear thermal current predicted in this paper characterizes the contribution from fermionic quasiparticles, like electrons. Similar intrinsic contributions is likely to arise in charge-neutral bosonic quasiparticles like magnons and phonons. Our density matrix based formalism can be possibly be adapted for these bosonic systems with some modifications. We leave this exploration for future study.}

\section{Conclusion}
\label{sec:SecIV}
To summarize, in this paper, we have developed the quantum kinetic framework for heat current induced by the temperature gradient in crystalline systems, and using this we study the second-order NL responses. We construct a definition of the thermal current in the quantum kinetic theory, which is consistent with the known semiclassical results in the linear response regime. We then extend this definition to calculate the second-order NL heat current.

In the second-order in temperature gradient, we find an intrinsic scattering time-independent heat current in addition to the extrinsic scattering time-dependent current. The intrinsic NL thermal current, as far as we know, has not been explored earlier and we predict it for the first time. In the dilute impurity limit, we show that the extrinsic current can be separated into linear and quadratic scattering time-dependent contributions. We found that the quadratic scattering time-dependent current corresponds to the thermal counterpart of the NLD current and the linear scattering time-dependent current corresponds to the thermal counterpart of the Berry curvature dipole current. These extrinsic current components were earlier obtained using the semiclassical theory. Our work provides the realization of these currents using the quantum kinetic theory.

We demonstrate that the intrinsic NL thermal current originates from the band geometric quantities of the system. Our symmetry analysis shows that to observe the intrinsic NL thermal current, both the inversion symmetry and the time-reversal symmetry must be broken. Parity-time reversal symmetric systems can be a  better platform to observe the intrinsic NL thermal Hall effect as the Berry curvature dipole-induced NL contribution vanishes.  

We employ our theory of NL thermal currents to demonstrate NL thermal conductivities in tilted massive Dirac system in the low-temperature limit. We find that within the constant relaxation time approximation, the NLD thermal current is independent of temperature, and it increases as the carrier density increases. The NLA Hall current peaks near the band edge, and it has a quadratic temperature dependence. The intrinsic contributions, on the other hand, have nontrivial inverse quadratic temperature dependence. The different temperature dependence, $\kappa^{\rm NLD} \propto T^0$, $\kappa^{\rm NLA} \propto T^2$, and $\kappa^{\rm NLI} \propto T^{-2}$ will be pivotal to extract out and identify the dominant contribution to the thermal current in low-temperature experiments. 


\section{Acknowledgement}
We acknowledge the Science and Engineering Research Board (SERB) and the Department of Science and Technology (DST) of the Government of India for financial support. H. V. thanks the MHRD, India, for funding through Prime Minister's Research Fellowship (PMRF). We sincerely thank Debottam Mondal for the useful discussion. 
\onecolumngrid
\appendix\label{Appendix} 
\section{\label{app_adiabatic}Calculation of the density matrix equation: Adiabatic switching-on approximation }
Here, we present a detailed calculation of the density matrix up to second order in temperature gradient $E_T$. Here, we use the following adiabatic switching-on approach, mostly used in the context of optics. In interaction picture, we can write
\be 
i\hbar \pdv{\tilde\rho(t)}{t} = [{\tilde{\m H}}_{E_T},{\tilde\rho}]~,
\ee 
where the tilde ($\sim$) sign over the operators represents that the operators are taken into the interaction picture, and ${\m H}_{E_T}$ conveys the perturbing Hamiltonian arising from the thermal perturbation. While writing the quantum-Liouville equation, we have used $-\frac{i}{\hbar} [{\m H}_{E_T}, \rho] = D_T(\rho)$. Thus, the form of the $N$-th order correction  to the density matrix can be obtained via
\be 
i\hbar \tilde\rho^{(N)}(t) = -\frac{i}{2} \int^t_{-\infty} dt' e^{\frac{i}{\hbar} {\m H}_0 t'} {\bm E}_T(t')\cdot \left[ \left\lbrace \m{H}_0 , \pdv{\rho^{(N-1)} (t')}{\kb} \right\rbrace - i[\bm{\m{R}_{\kb}},\lbrace \m{H}_0 , \rho^{(N-1)}(t') \rbrace]\right]e^{-\frac{i}{\hbar} {\m H}_0 t'}~.
\ee 
Here, we consider the thermal field of the form ${\bm E}_T (t) = {\bm E}_T e^{-i\omega t}$ and put $\omega = 0$ at the end for the dc counterpart. If we proceed with this, we will encounter integration of the form $\int^t_{-\infty} e^{-i\omega t'} dt'$. To solve this, we will use the adiabatic switching on of the thermal field. This can be achieved by the modified frequency $\omega + i \eta$. With this strategy, we solve the first-order density matrix as 
\bea 
&& \rho^{(1)}_{np}(t) = \frac{i}{2\hbar} \frac{1}{\omega_{np} - \omega - i \eta} \left[ \left\lbrace \m{H}_0 , \pdv{\rho^{(0)}}{k_b} \right\rbrace - i[\m{R}^b_{\kb},\lbrace \m{H}_0 , \rho^{(0)} \rbrace]\right]_{np}  E^b_T (t)~, \nn \\
&& \implies \rho^{(1)}_{np} = \frac{1}{2\hbar} \frac{1}{\omega_{np}  - i \eta} \left[ \left\lbrace \m{H}_0 , \pdv{\rho^{(0)}}{k_b} \right\rbrace - i[\m{R}^b_{\kb},\lbrace \m{H}_0 , \rho^{(0)} \rbrace] \right]_{np} E^b_T ~.
\eea 
In the second line of the above equation, we have used the dc limit ($i.e. \  \omega \to 0$). Likewise, we can calculate the second-order density matrix as 
\bea
&& \rho^{(2)}_{mp}(t) = \frac{i}{2\hbar} \frac{1}{\omega_{np} - 2\omega - 2i \eta} \left[ \left\lbrace \m{H}_0 , \pdv{\rho^{(1)}}{k_b} \right\rbrace - i[\m{R}^b_{\kb},\lbrace \m{H}_0 , \rho^{(1)} \rbrace]\right]_{np}  E^b_T (t)~, \nn \\ 
&& \implies \rho^{(2)}_{np} = \frac{i}{2\hbar} \frac{1}{\omega_{np}  - 2i \eta} \left[ \left\lbrace \m{H}_0 , \pdv{\rho^{(1)}}{k_b} \right\rbrace - i[\m{R}^b_{\kb},\lbrace \m{H}_0 , \rho^{(1)} \rbrace] \right]_{np} E^b_T ~.
\eea 
Physically, we can identify $\eta$ as the inverse of the relaxation time, {\it i.e.}, ~$\eta = 1/\tau$. Here, we notice that the prefactor of $\eta$ in the above equation depends upon the order of the calculated density matrix. For example, if we calculate $\rho^{(N)}$, then we will have a term like $(\omega_{np} - i N \eta)^{-1}$. This indicates that the higher-order component of the nonequilibrium density matrix originating from the multiphoton process decays faster. Another independent physical way to think of this is to associate $\eta$ with the broadening of the energy levels or the uncertainty in sampling the energy of the state. In this interpretation, it is natural to expect that multiphoton processes will have larger uncertainty in energy. If a single photon process samples two energy levels and has a frequency uncertainty of $\eta$ (or $1/\tau$), then a $N$- photon process will sample $2N$ energy levels is likely to have a frequency uncertainty of $N\eta$ or ($N/\tau$). 

Within the quantum kinetic theory framework, we can capture this  effect if we start with the quantum Liouville equation along with the relaxation time approximation in the following form, 
\be 
\pdv{\rho^{(N)}_{np}}{t} + \frac{i}{\hbar} [{\m H}_0,\rho^{(N)}]_{np} + \frac{\rho^{(N)}_{np}}{\tau/N} = [D_T(\rho^{(N-1)})]_{np}~.
\ee
Physically, this implies that the different orders of the density matrix (in powers of the external perturbation) relax with varying timescales of scattering, with $\rho^{(N)}$ relaxing with a timescale of $\tau/N$, as discussed above.

\section{Calculation of density matrix upto second order} \label{s1}
\subsection{First order density matrix}
Invoking the steady state of the density matrix, Eq.~\eqref{eqn:den3} can be written as 
\be\label{QKE} 
\frac{i}{\hbar}\left[ \m{H}_0,\rho^{(N)}\right]_{np} + \frac{\rho^{(N)}_{np}}{\tau/N} = \left[ D_T (\rho^{(N-1)})\right]_{np}.
\ee 
Using the commutation relation $[\m{H}_0,\rho^{(N)}]_{np} = (\tilde\e_n - \tilde\e_p ) \rho^{(N)}_{np}$ in the above equation, we calculated the general form of the $\rm{N}^{th}$ order density matrix to be 
\be\label{general_solution} 
\rho^{(N)}_{np} = -i\hbar g^{np}_N\left[ D_T (\rho^{(N-1)})\right]_{np}~,
\ee 
where we have defined $g_{np} = (\e_{np}-\frac{i\hbar}{\tau}N)^{-1}$ with $\e_{np} = (\e_{n} - \e_{p})$ being the interband energy gap.
By substituting $n = p $ in the above equation for $N = 1$ case, we calculate the diagonal elements of the first order density matrix ($\rho^{(1)}_{nn} \equiv \rho^{\rm{d}}_{nn} $). Using the identity $[D_T(\rho^{(0)})]_{nn} = \frac{1}{\hbar}\frac{\nabla_c T}{T} \tilde\e_n \partial_{c}f_{n}$ we obtain 
\be 
\rho^{\rm{d}}_{nn} = \frac{\tau}{\hbar}\frac{\nabla^c T}{T}  \tilde\e_n \partial_{c}f_{n} = - \frac{\tau}{\hbar} \tilde\e_n \partial_{c}f_{n} E^c_T.
\ee 
Furthermore, by using $[D_T(\rho^{(0)})]_{np} = \frac{i}{\hbar}\frac{\nabla^c T}{T}{\m{R}}^c_{np}(\tilde\e_n f_{n} - \tilde\e_p f_{p})$  for $n \neq p$ case, we calculate the off-diagonal elements of the first order density matrix ($\rho^{(1)}_{np} \equiv \rho^{\rm{o}}_{np}$) to be  
\bea 
\rho^{\rm{o}}_{np} = - \m{R}_{np}^{c}g^{np}_1 \xi_{np} E^c_T~,
\eea 
where we have introduced the notation $\xi_{np} = \tilde\e_n f_{n}-\tilde\e_p f_{p} $. With the help of the first-order density matrix, we will calculate the second-order density matrix in the next section.

%
\subsection{Second-order density matrix}
Now calculate the second order density matrix ($\rho^{(2)}$) by putting $N=2$ in Eq.~\eqref{general_solution}. Using a similar strategy as done for the first-order density matrix, we segregate the $\rho^{(2)}$ into the four parts as $\rho^{\rm{dd}},~\rho^{\rm{do}},~\rho^{\rm{od}},~{\rm{and}}~\rho^{\rm{oo}}$. Here, $\rho^{\rm{dd}}$ denotes the diagonal elements of the $\rho^{(2)}$ originating from $\rho^{\rm{d}}$, $\rho^{\rm{oo}}$ defines the off-diagonal elements of $\rho^{(2)}$ stemming from $\rho^{\rm{o}}$ and so on.
The diagonal elements of $\rho^{(2)}$ is calculated by using the identity
\be
[D_T(\rho^{(1)})]_{nn} = \frac{1}{2\hbar}\frac{\nabla^b T}{T} \left[ 2 \tilde\e_n \partial_{b} \rho^{\rm{d}}_{nn} + i \sum_{p \neq n } (\tilde\e_n + \tilde\e_p) ( \rho^{\rm{o}}_{np} {\m{R}}^{b}_{pn} - {\m{R}}^{b}_{np} \rho^{\rm{o}}_{pn}) \right].
\ee
From the first term of the above expression, we calculate the `dd' component of the second-order density matrix to be 
\be
\rho^{\rm{dd}}_{nn} = - \frac{\tau}{2 \hbar }\tilde\e_n \partial_b(\rho^{\rm{d}}_n)E^b_T =  \frac{\tau^2}{2 \hbar }\tilde\e_n \left[ v^n_b \partial_c f_n + \frac{\tilde\e_n}{\hbar} \partial_b\partial_c f_n\right] E^b_T E^c_T~.
\ee
The remaining part of $[D_T(\rho^{(1)})]_{nn}$ leads to the `do' component of the density matrix of the following form
\be 
 \rho^{\rm{do}}_{nn}  = \frac{i\tau}{4\hbar}\sum_{p \ne n} (\tilde{\e}_n + \tilde{\e}_p) \left(g^{np}_1\m{R}_{np}^{c}\m{R}_{pn}^{b} + g^{pn}_1\m{R}_{np}^{b}\m{R}_{pn}^{c}\right)\xi_{np} E^b_T E^c_T~.
\ee  
For the off-diagonal components of the NL density matrix ($n \neq p$ case), we use the identity
\be
[D_T(\rho^{(1)})]_{np}= \frac{1}{2\hbar}\frac{{\nabla}_bT }{T} \bigg[ (\tilde\e_n + \tilde\e_p)\partial_{b}\rho^{\rm{o}}_{np} + i \sum_{q} \bigg( (\tilde\e_n + \tilde\e_q)\rho^{(1)}_{nq}\m{R}^{b}_{qp} - (\tilde\e_q + \tilde\e_p)\m{R}^{b}_{nq}\rho^{(1)}_{qp} \bigg) \bigg].
\ee
The summation over $q$ in the second term inside the square bracket can be simplified by considering the three cases: $q = n \neq p $, $q = p \neq n $, and $q \neq n \neq p$. 
Using this mathematical trick, we simplify the above expression to the following form
\be 
[D_T(\rho^{(1)})]_{np} = \frac{1}{2\hbar}\frac{{\nabla}^bT }{T} \bigg[ 2 i (\tilde\e_n \rho^{\rm{d}}_{nn}  -  \tilde\e_p \rho^{\rm{d}}_{pp})\m{R}^b_{np} + (\tilde\e_n + \tilde\e_p)\m{D}^b_{np} \rho^{\rm{o}}_{np} + i\sum_{q \neq n \neq p } \left( (\tilde\e_n + \tilde\e_q)\rho^{\rm{o}}_{nq}\m{R}^{b}_{qp} - (\tilde\e_q + \tilde\e_p)\m{R}^{b}_{nq}\rho^{\rm{o}}_{qp} \right) \bigg]~,
\ee
where we have defined $\m{D}^b_{np} = \partial_b - i ( \m{R}^b_{nn} - \m{R}^b_{pp})$.
Using these results in Eq.~\eqref{general_solution}, the $\rho^{\rm{d}}$ dependent part of $[D_T(\rho^{(1)})]_{np}$ gives the `od' component of the off-diagonal density matrix of the following form
\be
\rho^{\rm{od}}_{np} = -i\hbar g^{mp}_2 \left[-\frac{i}{\hbar} E^b_T \m{R}^b_{np} (\tilde\e_n \rho^{\rm{d}}_{nn} - \tilde\e_p \rho^{\rm{d}}_{pp})  \right] =  \frac{\tau}{\hbar}g^{np}_2 \m{R}_{np}^{b}\left(\tilde{\e}_n ^2 \partial_{c}f_{n}-\tilde{\e}_p^2 \partial_{c}f_{p} \right)  E^b_T E^c_T~.
\ee  
The remaining $\rho^{\rm{o}}$ dependent part of $[D_T(\rho^{(1)})]_{np}$ gives the `oo' component of the off-diagonal density matrix having simplified expression
\bea
 \rho^{\rm{oo}}_{np}  &=& -\frac{i}{2}g^{np}_2 (\tilde{\e}_n + \tilde{\e}_p ) \m{D}^{b}_{np}\left(g^{np}_1\m{R}_{np}^{c}\xi_{np}\right) E^b_T E^c_T \nn \\ 
&+&  \frac{1}{ 2}g_2^{np} \sum_{q\neq n \neq p} \left[ g^{nq}_1\m{R}^{c}_{nq}\m{R}^{b}_{qp}(\tilde{\e}_n +\tilde{\e}_q )\xi_{nq}- g^{qp}_1\m{R}^{b}_{nq}\m{R}^{c}_{qp}(\tilde{\e}_q +  \tilde{\e}_p )\xi_{qp}\right]E^b_T E^c_T~.
\eea
The second term of the above equation is the multiband term. It becomes important for systems having three or more bands.

\section{Calculation of the linear thermal currents}\label{Appendix_LTC}
In this section, we will calculate the thermal current that is linear in the applied temperature gradient. Writing the first term of Eq.~\eqref{Th_C_def} in the band basis, we get
\be\label{LTC_1} 
{\rm{Tr}}\bigg[\frac{1}{2} \{{\m H}_0,{\bm v} \}\rho^{(1)}_{\rm D}\bigg]=\frac{1}{2} \sum_{n} \langle n \vert (\m{H}_0 \bm{v} + \bm{v}\m{H}_0) \rho^{(1)}_{\rm D} \vert n \rangle =  \sum_{n} \tilde\e_n  \bm{v}^{n} \rho^{\rm d}_{nn}~.
\ee
To keep track of its origin, we denote this current as ${\bm J}^{\rm d}$. After substituting the form of the $\rho^{\rm d}$, we get 
%
\be\label{LDTC} 
\bm{J}^{\rm d} = -\frac{\tau}{\hbar} \sum_{n, \kb } \tilde\e_n^2 \bm{v}^n \partial_c f_n  E^c_T~.
\ee 
This is the familiar expression of the  \textit{linear thermal Drude current} known in the literature~\cite{SSP}. The Drude thermal current is generally calculated within the semiclassical approach using the Boltzmann equation. Here, we calculate it from the quantum kinetic theory.
We denote the current originating from the second term Eq.~\eqref{Th_C_def} by ${\bm J}^{\rm o}$ as it stems from $\rho^{\rm o}$. In band-reduced form, it is calculated to be 
\be\label{LTC_2} 
{\bm J}^{\rm o} = \frac{1}{2} \sum_{p} \langle p \vert (\m{H}_0 \bm{v} + \bm{v}\m{H}_0) \rho^{(1)}_{\rm O} \vert p \rangle = \frac{1}{2} \sum_{n \neq p} (\tilde\e_n + \tilde\e_p) \bm{v}^{pn} \rho^{o}_{np}~.
\ee 
Using the off-diagonal elements of the velocity operator along an arbitrary spatial direction $a$ i.e. $v^{pn}_a = i\omega_{pn} \m{R}^a_{pn}$ and $\rho^{\rm o}_{np}$ in the above equation, we get 

\be  
J^{\rm o}_a =   \frac{i}{2\hbar} \sum_{n\neq p}  (\tilde\e_n + \tilde\e_p) \e_{np} g^{np}_1 \m{R}^a_{pn} \m{R}^c_{np}(\tilde\e_n f_{n} - \tilde\e_p f_{p}) E^c_T~.
\ee
In the dilute impurity limit, we have $g^{np}_1 \approx  1/\e_{np}$. Then the above equation is simplified further into the following form:
\be\label{j_o_linear}
J^{\rm o}_a   = -\frac{1}{2\hbar} \sum_{n\neq p}  \tilde\e_n (\tilde\e_n + \tilde\e_p) \Omega^{ac}_{np} f_n E^c_T~.
\ee 
%
%
Now, we focus on the third term of Eq.~\eqref{Th_C_def}, which gives thermal current contribution arising from the orbital magnetic moment (OMM) of the Bloch electrons. The particle magnetic moment of the n-th Bloch band is given by~\cite{Qian_Nu_PRL_2006}
\be 
{\bm m}_{N,n} = \frac{i}{2\hbar} \bra{{\bm \nabla}_{\kb} u^n_{\kb}} \times [ {\m H}_0 - \tilde\e_n]\ket{{\bm \nabla_{\kb}} u^n_{\kb}}~.
\ee 
Inserting the completeness relation, the above equation can be written in the following form,
\be\label{eq:omm_in_the bc_form}
\bal   
&{\bm m}_{N,n} = -\frac{i}{2\hbar} \sum_p (\e_n - \e_p) ({\bm {\m R}}_{np} \times {\bm {\m R}}_{pn} )~, \\
{\rm Thus,} & \ \epsilon_{abc} 
 m^a_{N,n} = -\frac{1}{2\hbar} \sum_p (\e_n - \e_p) \Omega^{bc}_{np}~.
\eal 
\ee 
Here, $\epsilon_{abc}$ is the third rank Levi-Civita tensor with $abc$ being the even cyclic permutation. Henceforth, we can compute the OMM contribution to the thermal current in the following way
\bea\label{J_OMM_1}
J^{\rm OMM}_a &=& {\rm{Tr}}\left[(\bm{E}_T \times \bm{m}_N)_a \m{H}_0 \rho_0 \right] = - {\rm{Tr}}\left[(\bm{m}_N \times \bm{E}_T)_a \m{H}_0 \rho_0 \right] =  -\sum_{n,\kb} \epsilon_{a b c}  m^b_{N,n} \tilde\e_n f_n E^c_T~, \nn \\ 
&=& -\frac{1}{2\hbar} \sum_{n,p,\kb} \tilde\e_n (\e_n - \e_p) \Omega^{ac}_{np} f_n E^c_T~.
\eea 
Lastly, focusing on the fourth term of Eq.~\eqref{Th_C_def}, which denotes the thermal current contribution from the Berry curvature-induced heat magnetization and is given by,
\bea\label{J_M_1} 
J^{\rm M}_a = 2{\rm Tr} (\bm{E}_T \times \bm{M}_{\bm \Omega})_a = - 2\sum_{n} \epsilon_{a  b c}  M^b_{{\bm \Omega},n} E^c_T~.
\eea 
Here, the Berry curvature-induced heat magnetization is given by $ M^b_{E,n} = \frac{1}{\hbar} \sum_{\kb} \zeta(\e_n)  \Omega^b_{n}$ where $\zeta (\e_n)$ is explicitly  given by
\be
\zeta(\e_n) = -\frac{1}{\beta^2} \left [  \frac{\pi^2}{6} - \frac{1}{2} \log^2 (1 + e^{-\beta(\e_n - \mu ) }) - {\rm Li}_2 (1-f_n)  \right]~.
\ee
Here Li$_2$ is the Polylogarithm function of order 2. From Eqs.~\eqref{j_o_linear}, \eqref{J_OMM_1}, and \eqref{J_M_1}, we note that all these contributions are nonzero only in the perpendicular direction of the applied temperature gradient. Consequently, they give rise to the novel anomalous thermal current response. All these anomalous thermal current components can be to have the following form
\bea 
J^{\rm Anm}_a &=& J^{\rm o}_{a} + J^{\rm OMM}_a + J^{\rm M}_a~, \nn \\ 
& = & \frac{1}{\hbar} \sum_n \tilde\e_n^2f_n \epsilon_{abc} \Omega^b_n E_T^c + \frac{(k_B T)^2}{ \hbar} \sum_n\left [  \frac{\pi^2}{3} -  \log^2 (1 + e^{-\beta(\e_n - \mu ) }) - 2 {\rm Li}_2 (1-f_n)  \right]\epsilon_{abc} \Omega^b_n E_T^c~. 
\eea 
Here, the superscript `Anm' stands for the `Anomalous.' Further, we can write the above equation in vector form as 
\be 
\bm{J}^{\rm Anm} = -\frac{(k_B T)^2}{\hbar} \bm{E}_T \times \int [d\kb] \sum_{n}  \bm{\Omega}_n \left[  \beta^2  \tilde\e_n^2 f_n +   \frac{\pi^2}{3} -  \log^2 (1 + e^{-\beta(\e_n - \mu )}) - 2 {\rm Li}_2 (1-f_n)  \right]~,
\ee 
%
This expression is identical to the anomalous thermal Hall (Righi-Leduc) currently known in the literature. It was earlier calculated in Refs.~[\onlinecite{Qian_PRL_2011,Shuichi_2011_PRB}] using the semiclassical equation of motion for thermal transport. 

\section{Calculation of the nonlinear thermal currents}\label{Appendix_NLTC}
In this section, we calculate the second-order $(N=2)$ thermal current. For the second-order thermal current, the definition is written from the Eq.~\eqref{Th_C_def_SC} as: 
\be\label{SOTC}
\bm{J}^{(2)}_a = {\rm{Tr}}\bigg[\frac{1}{2}\{\m{H}_0, v_a \}\rho^{(2)}_{\rm D}\bigg] + {\rm{Tr}}\bigg[\frac{1}{2}\{\m{H}_0, v_a \}\rho^{(2)}_{\rm O}\bigg] + {\rm{Tr}}\left[(\bm{E}_T \times \bm{m}_N)_a \m{H}_0 \rho^{(1)}_{\rm D}\right]~.
\ee
We begin with the first term of the above equation, which gives the NL heat current contributions from the diagonal elements of the second-order density matrix. The first term which we denote as $J_1^a$ can be written in the band reduced form as 
\be
J_1^a = {\rm{Tr}}\bigg[\frac{1}{2}\{\m{H}_0, v_a \}\rho^{(2)}_{\rm D}\bigg] = \frac{1}{2} \sum_{p,n} \langle p \vert \m{H}_0 v_a + v_a \m{H}_0 \vert n \rangle \langle n \vert \rho^{(2)}_{\rm D} \vert p \rangle = \sum_{n} \tilde\e_n   v^{n}_a \rho^{(2)}_{{\rm D},nn}~.
\ee
Here, we have used the fact that $\rho^{(2)}_{\rm D}$ is a diagonal matrix, hence it will be non-zero only for $p = n$. Since, we have $\rho^{(2)}_{\rm D} = \rho^{\rm dd} + \rho^{\rm do}$, these two sectors of the diagonal density matrix will give us two different thermal current contributions which we denote as $J^{\rm dd}_a$ and $J^{\rm do}_a$.
Using the form of $\rho^{\rm dd}_a$, we get 
\be 
J^{\rm dd }_a = \sum_{n} \tilde\e_n v^{n}_a \rho^{\rm dd }_{nn}= \frac{\tau^2}{2 \hbar^2 } \sum_{n,{\bm k}}  ( \hbar \tilde\e_n v_b^n \partial_c f_n  +  \tilde\e^2_n \partial_{b}\partial_c f_n)  \tilde\e_n v^n_{a} E^b_T E^c_T~.
\ee
Similarly, substituting the density matrix $\rho^{\rm do}$, we get another thermal current component
\bea
J^{\rm do}_a &=& \sum_{n} \tilde\e_n v^{nn}_a \rho^{\rm do }_{nn}= \frac{ i \tau}{4 \hbar}\sum_{n,p,\kb }^{p \ne n} (\tilde{\e}_n + \tilde{\e}_p) \left(g^{np}_1\m{R}_{np}^{c}\m{R}_{pn}^{b} + g^{pn}_1\m{R}_{np}^{b}\m{R}_{pn}^{c}\right) \tilde\e_n v^n_a (\tilde\e_n f_n - \tilde\e_p f_p) E^b_T E^c_T~, \nn \\ 
&=& \frac{i\tau}{4\hbar } \sum_{n,p, {\bm k}}^{p\ne n} g^{np}_1 \m{R}^b_{pn}\m{R}^c_{np} (\tilde\e_n + \tilde\e_p) (\tilde{\e}_n f_{n}- \tilde{\e}_p f_{p})\left(\tilde\e_n v^n_{a}- \tilde\e_p v^p_{a} \right) E^b_T E^c_T~.
\eea  
Now, we focus on the second term of Eq.~\eqref{SOTC}, which gives the thermal current contributions due to the off-diagonal components of the density matrix. We denote it by $J^a_2$ and simplify it as
\be 
J^a_2 = {\rm{Tr}}\bigg[\frac{1}{2}\{\m{H}_0, v_a \}\rho^{(2)}_{\rm O}\bigg]
=  \frac{1}{2} \sum_{p,n,\kb} \langle p \vert \m{H}_0 v_a + v_a \m{H}_0 \vert n \rangle \langle n \vert \rho^{(2)}_{\rm O} \vert p \rangle 
 = \frac{1}{2} \sum_{n,p,\kb}^{p \ne n} (\tilde\e_n + \tilde\e_p)  v^{pn}_a \rho^{(2)}_{{\rm O},np}~.
\ee
Note that the matrix element of the velocity operator is given by $v^{pn}_a = v^n_a \delta_{np} + i \omega_{pn} \m{R}^a_{pn}$. Since, in the main text we have considered $\rho^{(2)}_{\rm O} = \rho^{\rm od} + \rho^{\rm oo}$, therefore we will get two contributions. 
The thermal current contribution originating from $\rho^{\rm od}$ is given by
\be 
J^{\rm od}_a = \frac{1}{2} \sum_{n,p,\kb}^{p \ne n} (\tilde\e_n + \tilde\e_p)v^{pn}_a \rho^{\rm od}_{np} 
= \frac{i\tau}{2\hbar^2} \sum_{n,p, \kb}^{p \ne n} \e_{pn} g^{np}_2 ( \tilde\e_n + \tilde\e_p) \m{R}^a_{pn}\m{R}^b_{np}\left(\tilde\e_n^2 \partial_{c} f_{n} - \tilde\e_p^2 \partial_{c} f_{p} \right)  E^b_T E^c_T ~.
\ee
Similarly, we calculate the thermal current stemming from $\rho^{\rm oo}$ in the following way
\bea
  J^{\rm oo}_a &=& \frac{1}{2} \sum_{n\neq p} (\tilde\e_n + \tilde\e_p)v^{pn}_a \rho^{\rm oo}_{np}~, \nn \\ 
 &=& -\frac{1}{4\hbar } \sum_{n,p,\kb}^{p \ne n}  \e_{np} g^{np}_2 ( \tilde\e_n + \tilde\e_p )\m{R}^a_{pn}\bigg[ ( \tilde\e_n + \tilde\e_p ) \m{D}^b_{np}( g^{np}_1 \m{R}^c_{np} \xi_{np})  
\\
& +&  i \sum_{q\neq n \neq p} \left( g^{nq}_1 \m{R}^{c}_{nq}\m{R}^{b}_{qp}(\tilde{\e}_n +\tilde{\e}_q )\xi_{nq} - 
g^{qp}_1 \m{R}^{b}_{nq}\m{R}^{c}_{qp}(\tilde{\e}_q +  \tilde{\e}_p )\xi_{qp} \right)\bigg]E^b_T E^c_T~.
\eea 

Lastly, we focus on the third term of Eq.~\eqref{SOTC}, which gives the current contribution stemming from the orbital magnetic moment of the Bloch wave function. We denote this term by $J^{\rm OMM}$. It can be written in the following form,
\be
J^{\rm OMM}_a = \sum_{n} \langle n \vert (\epsilon_{a b l} E^b_T m^l_{N} \m{H}_0 \rho^{(1)}_{\rm D} \vert n \rangle
= \sum_{n}  \epsilon_{a b l} m^l_{N,n} \tilde\e_n \rho^{d}_{nn} E^b_T
= - \frac{\tau }{\hbar} \sum_{n,\kb} \epsilon_{a b l} m^l_{N,n} \tilde\e_n^2 \partial_c f_n E^b_T E^c_T~.
\ee  
NOw, inserting the Eq.~\eqref{eq:omm_in_the bc_form} in the above equation, we can write the OMM thermal current in the following form
\be 
J^{\rm OMM}_a = \frac{\tau }{2 \hbar^2} \sum_{n,p,\kb}  \tilde\e_n^2 (\tilde\e_n - \tilde\e_p) \Omega^{ab}_{np} \partial_c f_n E^b_T E^c_T~.
\ee 
This completes our calculation for all the thermal currents in the second-order response of the temperature gradient. We analyze these currents further in the main text.

\section{Intrinsic and extrinsic part of the scattering time-dependent factors}\label{the identities}
In this section, we present the mathematical steps to segregate the intrinsic and extrinsic parts of the thermal current. In the main text we have defined $g^{np}_N  = (\e_{np} - i\hbar N/\tau)^{-1}$. So, in the following section, we will show that the $\tau g^{np}_N$ can be segregated into the intrinsic and extrinsic parts. 
\be
\frac{\tau}{i\hbar} g^{np}_N = \frac{1}{\frac{i\hbar}{\tau}}\frac{1}{(\e_{np} - N \frac{i\hbar}{\tau})} = \frac{1}{\e_{np}}\left[ \frac{1}{\frac{i\hbar}{\tau}} + \frac{N}{\e_{np} -N\frac{i\hbar}{\tau}}\right]~.
\ee 
It is evident from the above equation that in the limit of $\tau \to \infty$, the first term of the above equation diverges, but the second term reduces to $N/\e_{np}^2$. Therefore, we use the following trick to write $\tau g^{np}_N$ into $\tau$ independent and dependent term: 
\be\label{identity_1}
\frac{\tau}{i\hbar} g^{np}_N = \frac{N}{\e_{np}^2} + \left( \frac{\tau}{i\hbar} g^{np}_N - \frac{N}{\e_{np}^2} \right) = \frac{N}{\e_{np}^2} + \frac{\tau}{i\hbar}\left( \frac{(\e_{np} + N\frac{i\hbar}{\tau})}{\e_{np}^2 + (\frac{N\hbar}{\tau})^2 } - \frac{N \frac{i\hbar}{\tau}}{\e_{np}^2}\right) = \frac{1}{\e_{np}^2} (N + \eta^{np}_{\tau,N})~,
\ee 
with $\eta^{np}_{\tau,N} = -i\tau\omega_{np} (1 -i (N/\tau\omega_{np})^3 )/(1+(N/\tau\omega_{np})^2)$, a dimensionless function of $\tau \omega_{np}$. So, the first term of the above equation is $\tau$ independent, while the second term is a function of $\tau$. In the case of dilute impurity limit where $\tau \omega_{np} \gg 1$ , we can approximate function $\eta^{np}_{\tau,N}(\tau \omega_{np})$  up to linear order in $\tau\omega_{np}$ as 
\bea 
\lim_{\tau\omega_{np} \to \infty} \eta^{np}_{\tau,N} (\tau\omega_{np} ) \approx -i\tau \omega_{np}~. 
\eea 
Similarly, we can show that in the $\tau \to \infty$, $g^{np}_N$ reduces to $1/{ \e_{np}}$. So, following the previous strategy, we calculated 
\be\label{identity_2} 
g^{np}_N = \frac{1}{\e_{np}} +\left( g^{np}_N -\frac{1}{\e_{np}} \right) = \frac{1}{\e_{np}}(1 - {\tilde \eta}^{\tau,N}_{np})~,
\ee 
with ${\tilde \eta}_{\tau,N}^{np} = i(N/\tau\omega_{np})(1 + i (N/\tau\omega_{np}))/(1 + (N/\tau\omega_{np})^2)$
being another dimensionless function of $\tau \omega_{np}$. It is straightforward to show that in the dilute impurity limit, the function ${\tilde \eta}_{\tau,N}^{np}$ goes to zero as 
\be 
\lim_{\tau \omega_{np} \to \infty}{\tilde \eta}_{\tau,N}^{np} (\tau \omega_{np}) \approx O\left( \frac{1}{\tau \omega_{np}}\right) \to 0~.
\ee 
Using these identities, we have separated the intrinsic and extrinsic contribution of the thermal current in the main text.
\section{Simplifiaction of the \texorpdfstring{$J^{oo}_{a, \rm{int}}$}{Joo}  current}\label{appendix_for_joo}
In the dilute impurity limit, the Eq.~\eqref{J_oo} can be written as 
\bea\label{j_oo_append}
J^{\rm oo}_{a,{\rm int}} &=&  -\frac{1}{4\hbar } \sum_{n,p,\kb}^{p \ne n}   ( \tilde\e_n + \tilde\e_p ) \m{R}^a_{pn} \bigg[  ( \tilde\e_n + \tilde\e_p ) \m{D}^b_{np} \left( \frac{1}{\e_{np}}\m{R}^c_{np} \xi_{np} \right) \nn \\ 
&& \ + i \sum_{q\neq n \neq p} \left( \frac{1}{\e_{nq}}\m{R}^{c}_{nq}\m{R}^{b}_{qp}(\tilde{\e}_n +\tilde{\e}_q )\xi_{nq} - 
\frac{1}{\e_{qp}} \m{R}^{b}_{nq}\m{R}^{c}_{qp}(\tilde{\e}_q +  \tilde{\e}_p )\xi_{qp} \right)\bigg] E^b_T E^c_T~.
\eea  
Now, focus on the first term of the above equation, which we denote by $J_{I}$ as  
\bea  
J_{I} &=& -\frac{1}{4\hbar } \sum_{n,p,\kb}^{p \ne n}   ( \tilde\e_n + \tilde\e_p )^2 \m{R}^a_{pn} \m{D}^b_{np} \left( \frac{1}{\e_{np}}\m{R}^c_{np} \xi_{np} \right)E^b_T E^c_T~. \nn \\
&=& -\frac{1}{4\hbar} \sum_{n,p,\kb}^{p\ne n }  ( \e_n + \e_p )^2 \m{R}^a_{pn} \left( \partial_b \left( \frac{1}{\e_{np}}\m{R}^c_{np} \xi_{np} \right) - i ({\m R}^b_{nn} - {\m R}^b_{pp})\left( \frac{1}{\e_{np}}\m{R}^c_{np} \xi_{np} \right) \right) E^b_T E^c_T~.
\eea 
With the help of algebraic manipulations, we can write the first term of the rounded bracket as 
\be   
\m{R}^a_{pn}  \partial_b \left( \frac{1}{\e_{np}}\m{R}^c_{np} \xi_{np} \right) 
=  \partial_b \left( \frac{1}{\e_{np}}  \m{R}^a_{pn} \m{R}^c_{np} \xi_{np} \right) 
 -  \frac{1}{\e_{np}}\m{R}^c_{np} \xi_{np} \partial_{b}\m{R}^a_{pn}
\ee  
Therefore, we can modify $J_{I}$ as 
\be 
J_{I} = -\frac{1}{4\hbar} \sum_{n,p,\kb} ( \e_n + \e_p )^2 \left[ \partial_b \left( \frac{1}{\e_{np}}  \m{R}^a_{pn} \m{R}^c_{np} \xi_{np} \right) 
- \frac{1}{\e_{np}}\m{R}^c_{np} \xi_{np} {\m D}^b_{pn} {\m R}^a_{pn} \right]E^b_T E^c_T~.
\ee 
Now, we modify the second part of eq.~\eqref{j_oo_append} by exchanging the dummy indices $i.e., \ q \leftrightarrow p \ and \ q \leftrightarrow n $. In this way, we can write the second part of Eq.~\eqref{j_oo_append} as 
\be 
J_{II} = -\frac{i}{4\hbar} \sum_{n,p,\kb}^{p \ne n } ( \tilde\e_n + \tilde\e_p )^2 {\m R}^c_{np} \frac{\xi_{np}}{\e_{np}} \sum_{q \ne (n,p)} ( {\m R}^a_{qn}{\m R}^b_{pq} - {\m R}^a_{pq}{\m R}^b_{qn})E^b_T E^c_T~.
\ee 
Thus, the intrinsic current $J^{oo}_{a,int}$ will become 
\be  
J^{oo}_{a,int} = -\frac{1}{4\hbar} \sum_{n,p,\kb}^{p \ne n} ( \tilde\e_n + \tilde\e_p )^2\left[\partial_b \left( \frac{1}{\e_{np}}  \m{R}^a_{pn} \m{R}^c_{np} \xi_{np} \right)  - \frac{{\m R}^c_{np} \xi_{np}}{\e_{np}}\left( {\m D}^b_{pn} {\m R}^a_{pn} - i \sum_{q \ne (n,p)} ( {\m R}^a_{qn}{\m R}^b_{pq} - {\m R}^a_{pq}{\m R}^b_{qn}) \right) \right]E^b_T E^c_T~.
\ee  
Now, using the sum rule~\cite{Watanbe_PRX_2021}, we can show that
\be  
{\m D}^a_{pn}{\m R}^b_{pn} - {\m D}^b_{pn}{\m R}^a_{pn} = -i \sum_{q \ne (n,p)} ( {\m R}^a_{qn}{\m R}^b_{pq} - {\m R}^a_{pq}{\m R}^b_{qn}))~.
\ee  
Thus, we can write
\be\label{j_oo_e8}  
J^{oo}_{a,int} = -\frac{1}{4\hbar} \sum_{n,p,\kb}^{p \ne n} ( \tilde\e_n + \tilde\e_p )^2\left[\partial_b \left( \frac{1}{\e_{np}}  \m{R}^a_{pn} \m{R}^c_{np} \xi_{np} \right)  - \frac{{\m R}^c_{np} \xi_{np}}{\e_{np}} {\m D}^a_{pn} {\m R}^b_{pn}\right]E^b_T E^c_T~.
\ee  
The permutation symmetry of indices $b$ and $c$ renders the non-trival part of the U(1) covariant derivative ${\m D}^a_{pn} {\m R}^b_{pn}$. This can be viewed by symmetrizing indices $b$ and $c$ followed by the exchange of dummy indices $n \leftrightarrow p$. With this manipulation, we can write the second term of the above equation as 
\bea 
J_2 &=& \frac{1}{4\hbar} \sum_{n,p,\kb} \frac{( \tilde\e_n + \tilde\e_p )^2}{\e_{np}} \xi_{np} {\m R}^c_{np} \partial_a {\m R}^b_{pn} E^b_T E^c_T~, \nn \\  
&=& \frac{1}{4\hbar} \sum_{n,p,\kb} \frac{( \tilde\e_n + \tilde\e_p )^2}{\e_{np}} \tilde\e_n f_n  \partial_a ({\m R}^c_{np}{\m R}^b_{pn})E^b_T E^c_T~, \nn \\ 
&=& \frac{1}{4\hbar} \sum_{n,p,\kb} \frac{( \tilde\e_n + \tilde\e_p )^2}{\e_{np}} \tilde\e_n f_n  \partial_a {\m G}^{bc}_{np} E^b_T E^c_T~.
\eea 
In obtaining this result, we have expanded the $\xi_{np}$ in the second line of the above equation and then used the exchange of dummy indices. Likewise, we transformed the first term of Eq.~\eqref{j_oo_e8} as 
\bea
J_1 &=& -\frac{1}{2\hbar} \sum_{n,p,\kb} ( \tilde\e_n + \tilde\e_p )^2 \partial_b \left( \frac{\tilde\e_n f_n }{\e_{np}} {\m G}^{ac}_{np} \right)E^b_T E^c_T~, \nn \\ 
&=& \frac{1}{2\hbar} \sum_{n,p,\kb}   \frac{\tilde\e_n f_n }{\e_{np}} {\m G}^{ac}_{np} \partial_b( \tilde\e_n + \tilde\e_p )^2 E^b_T E^c_T~,
\eea 
Finally, the simplified form of Eq.~\eqref{j_oo_append} has the following form:
\be 
J^{oo}_{a,{\rm int}} = \frac{1}{4\hbar} \sum_{n,p,\kb} ( \tilde\e_n + \tilde\e_p )\frac{\tilde\e_n f_n}{\e_{np}} \left[ ( \tilde\e_n + \tilde\e_p ) \partial_a {\m G}^{bc}_{np} + 4 {\m G}^{ac}_{np} \partial_b ( \tilde\e_n + \tilde\e_p )\right]E^b_T E^c_T~.
\ee 
%

\section{Exact expressions of extrinsic currents}
\label{ext_and_int}
In this section, we calculate the expressions of the extrinsic thermal currents. From Eq.~\eqref{NLD_therm} and Eq.~\eqref{OMM_current}, the explicit $\tau$ dependence of thermal currents $J^{\rm dd}$ and $J^{\rm OMM}$ are evident. However, for the remaining currents, it is not evident where the $\tau$ dependence is implicitly governed by either $\tau g^{np}_N$ or $g^{np}_N$. We will use identities in Eq.~\eqref{identity_1} and Eq.~\eqref{identity_2}, to extract the $\tau$ dependence of different thermal currents. Using this approach, we calculate the extrinsic part of $J^{\rm do}_a$ with the help of identity of Eq.~\eqref{identity_1} in the following form,
\be\label{J_do_ext_1}
J^{\rm{do}}_{a,\rm{ext}} = -\frac{1}{4 } \sum_{n,p, {\bm k}}^{p\ne n} \frac{\eta^{np}_{\tau,1}}{\e_{np}^2}\m{R}^b_{pn}\m{R}^c_{np} (\tilde\e_n + \tilde\e_p) (\tilde{\e}_n f_{n}- \tilde{\e}_p f_{p}) \left(\tilde\e_n v^n_{a}- \tilde\e_p v^p_{a} \right) E^b_T E^c_T~.
\ee
Further, we can split $\eta^{\tau}_{np}$ into symmetric and anti-symmetric parts under exchange of $n \leftrightarrow p$. Using this idea in the above equation, we observed that the anti-symmetric part cancels out, and the symmetric part gives a finite contribution. So, the nonvanishing contribution of Eq.~\eqref{J_do_ext_1} is 
\be\label{J_do_ext}
J^{\rm{do}}_{a,\rm{ext}} = \frac{1}{2 } \sum_{n,p,\kb} \m{G}^{bc}_{np} \frac{\tilde\e_n (\tilde\e_n + \tilde\e_p)}{ \e_{np}^2 (1 + \tau^2 \omega_{np}^2)}  \left(\tilde{\e}_n v^n_{a}-\tilde{\e}_p v^p_{a}\right) f_n E^b_T E^c_T~.
\ee
We further notice that in the dilute impurity limit, $J^{\rm{do}}_{a,\rm{ext}} \to 0$.

Similarly, we derived the extrinsic part of the $J^{\rm od}_a$ to be
\be\label{Ext_od}
J^{\rm{od}}_{a,\rm{ext}} = -\frac{1}{ \hbar} \sum_{n,p} \int [d\kb]  \frac{\tilde\e^2_n(\tilde\e_n+ \tilde\e_p) }{\e_{np} (1 + (\tau \omega_{np}/2)^2)} \left( 2 \m{G}^{ab}_{np} - \left(\frac{\tau \omega_{np}}{2}\right)^3 \Omega^{ab}_{np} \right) \partial_c f_n E^b_T E^c_T~.
\ee
In the dilute impurity limit, this expression reduces to Eq.~\eqref{J_od_extrinsic}.

Lastly, with the help of identity Eq.~\eqref{identity_2}, we calculated the extrinsic part of the $J^{\rm oo}_a$ to be
\bea\label{J_oo_ext}
 && J^{\rm oo}_{a,ext} =  -\frac{1}{4\hbar } \sum_{n,p,\kb}^{p \ne n}   \tilde{\eta}^{np}_{\tau,2} ( \tilde\e_n + \tilde\e_p )\m{R}^a_{pn}\bigg[ ( \tilde\e_n + \tilde\e_p ) \m{D}^b_{np}( \tilde{\eta}^{np}_{\tau,1} \m{R}^c_{np} \frac{\xi_{np}}{\e_{ng^{np}_1}})  
\\
& +&  i \sum_{q\neq n \neq p} \left( \tilde{\eta}^{nq}_{\tau,1} \m{R}^{c}_{nq}\m{R}^{b}_{qp}(\tilde{\e}_n +\tilde{\e}_q )\frac{\xi_{nq}}{\e_{nq}} - 
\tilde{\eta}^{qp}_{\tau,1} \m{R}^{b}_{nq}\m{R}^{c}_{qp}(\tilde{\e}_q +  \tilde{\e}_p )\frac{\xi_{qp}}{\e_{qp}} \right)\bigg]E^b_T E^c_T~.
\eea 

\section{Comparison between nonlinear electrical and thermal currents: Validity of Wiedemann-Franz law}
In the linear response of the external field, the electrical conductivity ($\sigma_{ab}$) and the thermal conductivity ($\kappa_{ab}$) are related by the Wiedemann-Franz (WF) law: $\kappa/\sigma = L T$, {where $L = \frac{1}{3} \left( \frac{\pi k_B}{e} \right)^2$ is the Lorentz number~\cite{lorentz_number}}. However, in the NL response regime, this relation does not hold. In this section, we compare the NL electrical and thermal conductivity and comment on the relations between the NL charge and thermal conductivity. We start with the linear scattering time-dependent component of the currents. The charge current due to the electric field and the thermal current due to the temperature gradient is given by
\bea
 j_{a} (\tau) &=& - \frac{\tau e^3}{\hbar^2} \sum_{n,p,\kb} \Omega_{np}^{ab} \partial_c f_n E_b E_c~, \\ 
J_{a} (\tau) &=&  \frac{\tau}{\hbar^2} \sum _{n,p,\kb}  \Omega_{np}^{ab}\tilde\e^3_n \partial_c f_n E^b_T E^c_T~.
\eea 
These extrinsic parts of the currents are related to each other by the modified Wiedemann-Franz law for NL response~\cite{Nandy20}. Mathematically, it can be represented by
\be 
\kappa_{abc} = \frac{14}{15} e L_0^2 T^2 \pdv{\chi_{abc}(\e)}{\e}\bigg\vert_{\mu}~,
\ee 
where we have defined, $L_0 =\left( \frac{\pi k_B}{e} \right)^2$ and $\chi_{abc} = \frac{ \tau e^3}{2\hbar} \int [d\kb] \epsilon_{l a b} \Omega^l_n v^n_c$ is the zero temperature NL anomalous Hall conductivity. Now we focus on the quadratic scattering time-dependent NL current. The charge current due to the electric field and the thermal current due to the temperature gradient is given by
\bea
 j^{\rm{dd}}_a &=& -\frac{\tau^2 e^3}{\hbar^2}\sum_{n, {\bm k}}  v^n_a \partial_b \partial_c f_n E_b E_c~,  
 \\ 
 J^{\rm{dd}}_{a}  &=& \frac{\tau^2}{2\hbar^2 } \sum_{n, {\bm k}}  \tilde\e_n v^n_{a} ( \hbar \tilde\e_n v_b^n \partial_c f_n  +  \tilde\e^2_n \partial_{b} \partial_{c} f_n)  E^b_T E^c_T~.
\eea
From the above expressions, one can infer that the NL Drude currents do not satisfy the modified Wiedemann-Franz law. Similarly, the intrinsic currents also do not satisfy any general relation.

\twocolumngrid
\bibliography{QKT}
\end{document}